\documentclass[aps,prx,twocolumn,superscriptaddress,floatfix]{revtex4-2} 
\usepackage[utf8]{inputenc}		
\usepackage[english]{babel}
\usepackage{breakcites}
\usepackage{lmodern}
\usepackage[normalem]{ulem}

\usepackage{ae,aecompl}
\usepackage{subfigure}
\usepackage{latexsym,stmaryrd}
\usepackage{graphicx}
\usepackage[font={normalsize},labelsep=endash]{caption}
\usepackage[hidelinks,linkcolor=blue,colorlinks=true,urlcolor=blue,citecolor=blue]{hyperref}
\usepackage{braket}
\usepackage{amsmath}
\usepackage{upgreek}
\usepackage{cleveref}
\usepackage{url}
\usepackage{amssymb}

\newcommand{\pos}{x_\mathrm{\ell}}
\newcommand{\pin}{P_\mathrm{in}}
\newcommand{\Gm}{\Gamma_\mathrm{m}}
\newcommand{\Wm}{\Omega_\mathrm{m}}
\newcommand{\um}{\upmu\textrm{m}}
\newcommand{\as}{a_{\textrm{s}}}
\newcommand{\ac}{a_{\textrm{c}}}

\begin{document}
\renewcommand{\figurename}{FIG.}
\crefname{section}{Sec.}{Secs.}


\title{Optomechanical generation of coherent GHz vibrations in a phononic waveguide}


\author{Guilhem Madiot}
\thanks{These authors contributed equally to this work}
\affiliation{Catalan Institute of Nanoscience and Nanotechnology (ICN2), Campus UAB, Bellaterra, 08193 Barcelona, Spain}

\author{Ryan C. Ng}
\thanks{These authors contributed equally to this work}
\affiliation{Catalan Institute of Nanoscience and Nanotechnology (ICN2), Campus UAB, Bellaterra, 08193 Barcelona, Spain}
\affiliation{Corresponding authors: ryan.ng@icn2.cat and david.garcia@icn2.cat}

\author{Guillermo Arregui}
\affiliation{DTU Electro, Department of Electrical and Photonics Engineering, Technical University of Denmark, Ørsteds Plads 343, DK-2800 Kgs. Lyngby, Denmark}

\author{Omar Florez}
\affiliation{Catalan Institute of Nanoscience and Nanotechnology (ICN2), Campus UAB, Bellaterra, 08193 Barcelona, Spain}
\affiliation{Dept. de F\'{i}sica, Universitat Autonoma de Barcelona, 08193 Bellaterra, Spain}

\author{Marcus Albrechtsen}
\affiliation{DTU Electro, Department of Electrical and Photonics Engineering, Technical University of Denmark, Ørsteds Plads 343, DK-2800 Kgs. Lyngby, Denmark}

\author{S{\o}ren Stobbe}
\affiliation{DTU Electro, Department of Electrical and Photonics Engineering, Technical University of Denmark, Ørsteds Plads 343, DK-2800 Kgs. Lyngby, Denmark}
\affiliation{NanoPhoton - Center for Nanophotonics, Technical University of Denmark, {\O}rsteds Plads 345A, DK-2800 Kgs.\ Lyngby, Denmark}

\author{Pedro D. Garc\'{i}a}
\affiliation{Catalan Institute of Nanoscience and Nanotechnology (ICN2), Campus UAB, Bellaterra, 08193 Barcelona, Spain}
\affiliation{Corresponding authors: ryan.ng@icn2.cat and david.garcia@icn2.cat}

\author{Clivia M. Sotomayor-Torres}
\affiliation{Catalan Institute of Nanoscience and Nanotechnology (ICN2), Campus UAB, Bellaterra, 08193 Barcelona, Spain}
\affiliation{ICREA - Instituci\'o Catalana de Recerca i Estudis Avan\c{c}ats, 08010 Barcelona, Spain}

\begin{abstract}
Nanophononics has the potential for information transfer, in an analogous manner to its photonic and electronic counterparts. The adoption of phononic systems has been limited, due to difficulties associated with the generation, manipulation, and detection of phonons, especially at GHz frequencies. Existing techniques often require piezoelectric materials with an external radiofrequency excitation that are not readily integrated into existing CMOS infrastructures, while non-piezoelectric demonstrations have been inefficient. In this work, we explore the optomechanical generation of coherent phonons in a suspended 2D silicon phononic crystal cavity with a guided mode around 6.8 GHz. By incorporating an air-slot into this cavity, we turn the phononic waveguide into an optomechanical platform that exploits localized photonic modes resulting from inherent fabrication imperfections for the transduction of mechanics. Such a platform exhibits very fine control of phonons using light, and is capable of coherent self-sustained phonon generation via mechanical lasing around 6.8 GHz.\ The ability to generate high frequency coherent mechanical vibrations within such a simple 2D CMOS-compatible system could be a first step towards the development of sources in phononic circuitry and the coherent manipulation of other solid-state properties.\
\end{abstract}

\maketitle

\section{Introduction}

Phononics has received less attention relative to its photonic and electronic counterparts, largely due to challenges associated with generating, transporting, manipulating, and detecting phonons, despite their potential as carriers of information signals \cite{Wang2007,Li2012,Sklan2015,Ng2022arXiv,Zivari2021arXiv,Zivari2022arXiv}.\ The utility of phononics lies in its ability to operate in the MHz and GHz regimes, which connects the frequency regimes of electronics and optics \cite{Hatanaka2014,Laer2015,Fu2019}.\ 
In an analogous manner to photonic crystals, the dispersion relation and mode profiles of acoustic phonons can be engineered by making structures with periodic elastic properties.\ In planar structures, careful design coupled with improvements in nanofabrication have enabled the realization of mechanical band gaps up the GHz frequency range \cite{su2010realization,soliman2010phononic}, while the introduction of linear defects into these crystals permits phonon waveguiding \cite{Pourabolghasem2018,Dehghannasiri2018,Florez2022}.\ However, the excitation of phonons in phononic circuits still remains a challenge, with existing approaches relying on electromechanical actuation in piezoelectric materials \cite{Fuhrmann2011, Mahboob2013, Balram2016, Korovin2019,Kuznetsov2021}.\ While such an approach does in theory offer the technical advantages of ultra-compact on-chip integrability with low power consumption, piezoelectric solutions always rely on an external RF signal that is not as readily integrated.\ Furthermore, these approaches are lacking in their ability to control and manipulate phonons and are restricted to a limited bandwidth that is often set by the specific interdigitated transducer design used for excitation.\ Surface acoustic wave excitation in a non-piezoelectric platform has recently been shown via thermo-elastic modulation \cite{munk2019surface} although this process remains quite inefficient.\ As an alternative, cavity optomechanical platforms are able to finely control the phonon signal via radiation pressure forces while relying on an integrated photonic circuit rather than an external RF source, circumventing these aforementioned problems.\ Within these platforms, self-sustained mechanical oscillations driven by light, otherwise known as phonon lasing \cite{Grudinin2010,Burek:16,NavarroUrrios2015,ghorbel2019optomechanical,mercade2021floquet}, can potentially be used to generate a phonon signal directly within a phononic circuit \cite{Fang2016,Fu2019,Guo2019,Mayor2021}.\ This is typically achieved by side-coupling an optomechanical cavity to another phononic waveguide and requires proper matching of the respective mechanical bandwidths and mode profiles \cite{Patel2018,zivari2022non}. A solution that does not require interdigitated transducers or more elaborate optomechanical architectures is thus an important building block for phononic circuitry.\\

\begin{figure*}
\centering
\includegraphics[trim={0.3cm 1cm 0.15cm 0.8cm},clip]{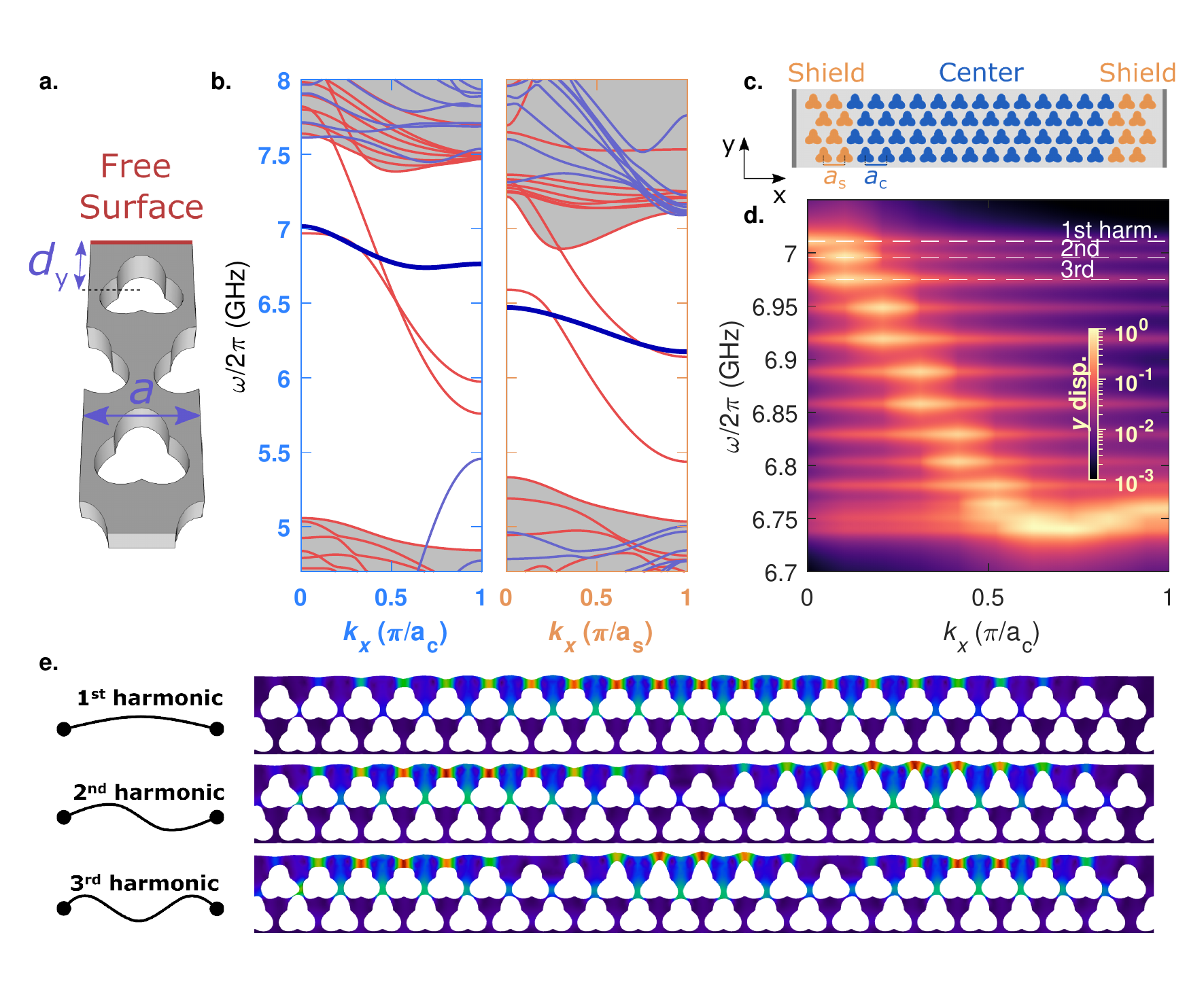}
\caption{\textbf{Acoustic waveguide design on a two-dimensional shamrock phononic crystal.}  (a) Supercell of the waveguide geometry, with the free boundary at the silicon-air interface highlighted in red.\ (b) Associated mechanical band structure for the center and shield regions showing the symmetric (blue) and antisymmetric (red) modes. As the modes which are antisymmetric with respect to the \textit{z} mid-plane are optomechanically dark, the guided mechanical mode in the center region is confined by the gap of the shield region. (c) Schematic illustration of the phononic waveguide incorporating a center cavity region ($n_a=0.23\ac$, $n_b=0.30\ac$, $\ac=500$ nm) and shield region with slightly larger period along the \textit{x} direction ($\as=560$ nm) to confine the mechanical mode.\ The schematic is not to scale.\ (d) In-plane displacement spectrum plotted as a function of the k-vector. This simulation is made accounting for the full structure within one simulation (i.e., both the center and shield regions together), and the acoustic Fabry-Pérot fringes can be observed. (e) Displacement mode profiles for the full structure showing the acoustic confinement within the center region for the first three harmonics.}\
\label{Fig1}
\end{figure*}




Here, we present a 2D silicon phononic crystal waveguide sustaining an acoustic guided mode around 6.8 GHz and exploit optomechanical interactions to transduce and generate coherent acoustic phonons directly within the phononic waveguide itself.\ 
To realize this, an optomechanical interface is obtained by bringing together two identical phononic crystal cavities with mirror symmetries separated by an air gap defined by their interfaces.
This results in a slot photonic-crystal waveguide that allows for the optomechanical readout and transduction of mechanics by employing naturally-emerging disorder-induced localized optical cavities.\ 
Confinement of the mechanical waveguide modes induced by reflections at the ends of the waveguide result in the observation of acoustic Fabry-Pérot fringes, with the associated strong optomechanical interaction leading to dynamical back-action and subsequent mechanical lasing of certain mechanical modes.\ From a device perspective, the ability to observe mechanical lasing and coherently generate phonons in such a platform is a first step towards high-throughput coupling to other mode-matched acoustic waveguides.\

\section{Phononic waveguide design}
\label{design}

We design a phononic crystal (PnC) waveguide based on a 220 nm silicon membrane where \textit{shamrock}-shaped holes \cite{Wen:08} tiled in a triangular lattice are etched, as schematically illustrated in \Cref{Fig1}a.\ The shamrocks are the result of three overlapping elliptical holes, with each ellipse defined by its short and long axes given by $n_{\textrm{a}} = 0.23\ac$ and $n_{\textrm{b}} = 0.30\ac$, respectively, where $\ac = 500$ nm is the PnC lattice period.\ Such a PnC opens a large phononic band gap centered around 6.2 GHz, which enables the creation of guided phononic waveguides through line defects. We achieve this by terminating a crystal with a straight free-surface, as indicated in red in \Cref{Fig1}a, resulting in an acoustic waveguide that can be thought of as a silicon "wire" with edges defined by air and by the shamrock PnC \cite{Patel2018}. The row of shamrocks closest to the free surface is shifted away from the edge by $d_{\textrm{y}}\sqrt{3}\ac/2$, where $d_{\textrm{y}}=0.1155$.\ Subsequent rows along the $y$ direction are kept at their nominal positions in the crystal lattice.\ The distance between the free surface and the center of the first row of shamrocks sets the width of the waveguide region, $d_\textrm{wg}$, and controls the number of supported modes. We calculate the band structure of the waveguides using finite-element-method (FEM) simulations and show two examples of this in \Cref{Fig1}b. In these band structures, the bands are associated to $z$-symmetric or $z$-antisymmetric modes (i.e., symmetric or antisymmetric with respect to the center mid-plane, shown in blue and red, respectively). Due to the optomechanical transduction described later, we focus only on the $z$-symmetric modes. By choosing $d_\textrm{wg} = 190$ nm, the waveguide possesses a single propagating $z$-symmetric guided mode, highlighted by a thick blue line in \Cref{Fig1}b for two geometries with different horizontal pitches, $a_x$, of 500 nm (light blue border) and 560 nm (yellow border), both with the same shamrock dimensions. The combination of the above parameters allows for the band of interest to be flattened at the Brillouin zone edge as well as to be shifted away from the band gap edge to prevent coupling to bulk modes.\

We explore the phononic waveguide with band structure given in the left of \Cref{Fig1}b (light blue border), where both the horizontal and vertical pitch coincide. To improve the temporal confinement of the acoustic modes supported, we discretize the waveguide spectrum by creating a long acoustic cavity in the direction of the waveguide ($x$).\ This cavity is formed by sandwiching a center region with horizontal pitch $a_{\textrm{c}}$ = 500 nm between two shield regions with horizontal pitch $a_{\textrm{s}}$ = 560 nm, respectively shown with blue and yellow in the schematic of \Cref{Fig1}c.\ While both the center and shield regions possess a guided mechanical mode within the band gap as shown in \Cref{Fig1}b, the greater horizontal period of the shield region shifts the symmetric guided mode to a lower frequency and prevents coupling between the modes of each region. This results in an acoustic Fabry-Pérot cavity for the mode in the central region, which significantly increases the mechanical quality factor, $Q_{\text{m}}$, of the acoustic phonons \cite{GomisBresco2014}.\ We study structures with varying center region lengths $L$, while the shield region is set to have fixed 8 unit cells along the $x$ direction in all cases. \Cref{Fig1}d shows the reconstructed waveguide band obtained after Fourier-transforming the in-plane $y$-displacement along the free surface for the modes supported by a structure of length $L$=20$a_{\textrm{c}}$. An identical mechanical Q-factor of $Q_m=700$ is assumed for all modes regardless of the radiation-limited value obtained through simulation, in accordance with the experimental observations that follow. The shield structure effectively generates a long cavity that inherits the dispersion properties of the underlying guided mode in the center region (\Cref{Fig1}b, left). This is also observed when the mode profiles are explored in real space, shown via the first three harmonics in \Cref{Fig1}e. The displacement profile results from the waveguide mode and is thus concentrated at the free surface, dominated by an in-plane component along the \textit{y} direction, and decays quickly within the crystal away from this free surface. Along the waveguide axis, the modes have an increasing number of nodes and anti-nodes, mimicking a doubly-clamped string. This simulation is realized with a short cavity for computational efficiency, but the same physical picture also holds in longer structures. Unlike the simulations shown in \Cref{Fig1}d and \Cref{Fig1}e, a realistic waveguide is subject to fabrication imperfections such as sidewall roughness. Such disorder ultimately transforms the spectro-spatial acoustic properties in the slow sound region (around the $\Gamma$ point) when the length of the waveguide overcomes the acoustic localization length \cite{PhysRevB.95.115129}, a feature that we recover through FEM simulations of long waveguides subject to a minimal positional disorder model. While this same effect is exploited in the optical domain in the following section, experimentally distinguishing between Fabry-Pérot cavity modes and disorder-induced mode localization is a complicated task that is outside the present scope.

\section{Optomechanical cavity}
\label{opt_modes}

\begin{figure*}[!ht]
\centering
\includegraphics[trim={0.2cm 0.8cm 0.15cm 1.6cm},clip]{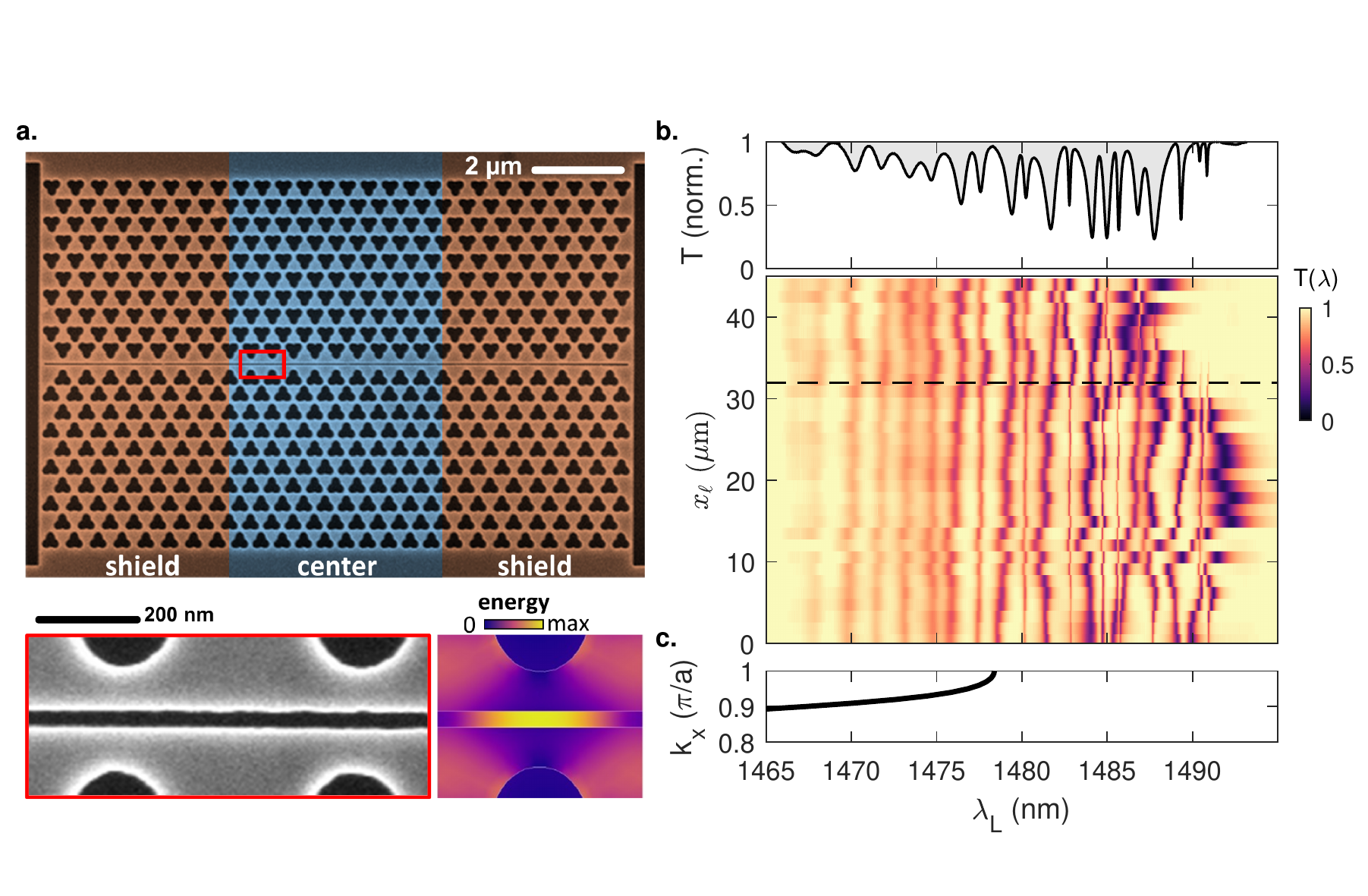}
\caption{ \textbf{Optical characterization of a slotted two-dimensional optomechanical waveguide.} (a) False-color SEM image of the shamrock optomechanical cavity waveguide formed by bringing two PnC waveguides of opposite symmetry from \Cref{Fig1}a together, separated by an air-slot region. The air-slot enables the excitation of mechanical modes via localized optical modes when a tapered fiber loop is brought into contact with the waveguide.\ A higher magnification SEM image focused on the center slot is highlighted in red.\ The electromagnetic energy density from FEM simulations shows the slot-confined light within the thin air gap between both PnC's (bottom right).\ (b) A representative optical transmission spectrum (top) is taken with a tapered fiber loop, where dips in transmission indicate coupling to a resonant optical mode in the structure. A spatially varying spectral map (bottom) shows the transmission spectrum (color scale) as the fiber loop is moved along the length of the waveguide.\ The dotted line is the position at which a representative optical transmission spectrum is taken.\ In the spectral map, at lower wavelengths, the periodic dips in transmission indicate the extended Fabry-Pérot modes, while the modes at higher wavelengths, which are more irregular in wavelength, are localized modes.\ (c) Optical band structure showing the guided optical mode alongside the spectral map and representative spectrum, where the guided mode flattens at the Brillouin zone edge, indicating a high group index, or low group velocity and subsequent mode localization.}
\label{Fig2}
\end{figure*}

The shamrock crystal proposed for the phononic waveguide also supports a photonic band gap \cite{ArreguiPRB2018}. The acoustic design sustains the guided mechanical mode of interest at a silicon-air free surface. However, this free surface is normally an inconvenient location for a photonic waveguide mode as air does not confine light, which complicates the direct optomechanical transduction of motion. Instead, we can exploit this otherwise disadvantageous nature of the mechanical waveguide by adding a mirror-symmetric version of the waveguide a distance of 40 nm away from the original, which leads to two independent mechanical membranes separated by a center air-slot region.\ 
The resulting design is shown in the false-color SEM image of \Cref{Fig2}a.\ Such an approach leads to a slot photonic-crystal waveguide for the optics. This allows the electromagnetic field to be locally enhanced in the slot while guiding light along its length, creating a cavity-optomechanical system that enables light to couple to in-plane mechanical motion \cite{Arregui2021arXiv}. A FEM simulation of the electromagnetic energy density is shown in the bottom right of \Cref{Fig2}a and additional details about the slot-guided mode and the associated photonic band structure are provided in \Cref{SI:Photonics}.\\

The suspended silicon structures are fabricated on a commercial silicon-on-insulator wafer following the fabrication process flow described in \cref{fabrication}. The resulting structures are optically characterized with a tapered fiber loop that is placed in contact with the structure along the waveguide axis, allowing light that is travelling in the fiber to couple evanescently into resonant optical modes of the optomechanical cavity structure \cite{GomisBresco2014}.\ The mechanical shield region plays an analogous role for the slot-guided optical mode, making the structure a very long optical cavity-waveguide  (see \cref{SI:Photonics}). A representative optical spectrum taken at a central position along a waveguide of length $L_\textrm{total}\sim$84 $\um$, is shown in the top panel of \Cref{Fig2}b, in which resonant optical modes manifest as spectral transmission dips. To understand the spatial structure of the observed optical modes, we acquire transmission spectra at different fiber loop positions, denoted as $\pos$, along the waveguide axis. Spectra are acquired in steps of $\sim2$ $\um$ and $\pos=0$ and $\pos=75$ $\um$ denote the two bounds of the central region of the waveguide.
The resulting spatially varying spectral map is shown in \Cref{Fig2}b (bottom).\ The black dotted line indicates the specific position at which the representative transmission spectrum shown in the top panel of \Cref{Fig2}b is taken. At the shorter wavelength end of the spectrum, periodic dips in transmission are observed, which are extended Fabry-Pérot modes spanning the entire waveguide and result from reflection at the waveguide edges, similar to the mechanical modes shown in \Cref{Fig1}e.\ At higher wavelengths the spectral location of the modes is less regular and the modes are spatially localized to specific sections of the waveguide in a manner akin to Anderson localization \cite{PhysRev.109.1492, Arregui2021arXiv}. This results from localization of the light field due to multiple scattering off of nanoscale fabrication imperfections such as sidewall roughness resulting from etching at the silicon-air interface (see SEM inset of the zoom-in of the slot in \Cref{Fig2}a) \cite{LeThomas2009}. Any further mention of optical localization in this work implicitly refers to disorder-induced localization. This localization occurs at longer wavelengths where the slot-guided light is slowed down, as indicated by the slot-mode dispersion in the central region (\Cref{Fig2}c) where the band flattens near the Brillouin zone edge.\ In this regime, light-matter interactions are enhanced and the system becomes increasingly sensitive to any nanoscale imperfections leading to localized optical cavities with a mode volume that decreases with increasing group index \cite{Sapienza2010}. At even longer wavelengths, after approximately $\lambda$ = 1488 nm, the cut off of the guided mode is reached and near-unity transmission is observed where there is no available optical mode to couple into.\\

\begin{figure*}
\centering
\includegraphics[trim={0cm 0cm 0.0cm 0.1cm},clip]{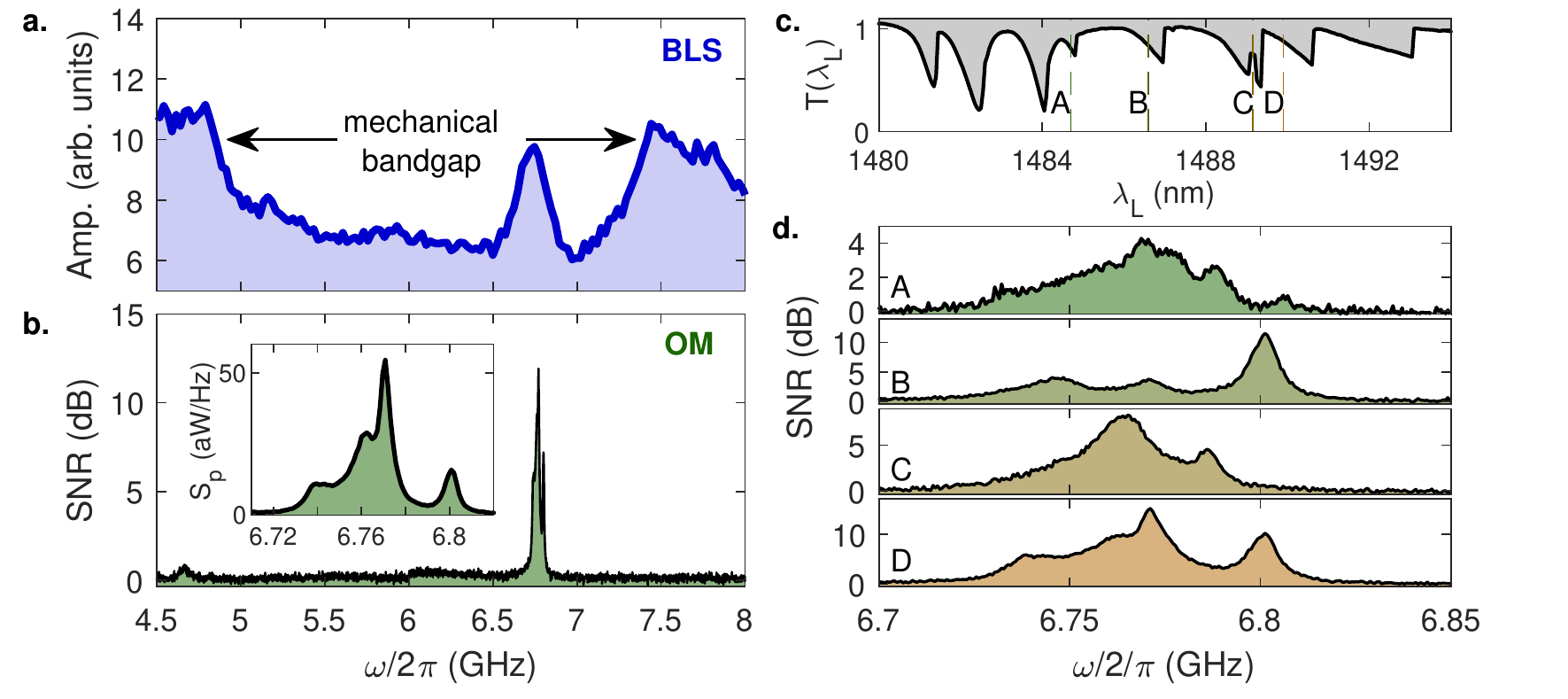}
\caption{\textbf{GHz mechanical spectroscopy of an acoustic cavity-waveguide.} (a) Brillouin light scattering (BLS) spectroscopy measurement in which the GHz mechanical guided mode can be observed within the mechanical band gap.\ (b) The same mechanical mode is probed with the tapered fiber loop over the same frequency range as in the BLS measurement to allow for direct comparison.\ The inset shows a magnified spectrum of the same mode, allowing for the different peaks to be more clearly observed. (c) Optical transmission spectrum when probing with a tapered fiber loop showing localized optical-modes that are slightly shifted by the thermo-optic effect. The mechanical spectra that result from driving 4 of these optical modes (as indicated by dashed lines in (c)) are shown in (d).\ }
\label{Fig3}
\end{figure*}

While inherent fabrication imperfections such as these are commonly considered a hindrance, here we exploit them to localize the optical field into optical cavities that do not require deliberate adiabatic tuning of the geometry. The localized optical modes supported by this structure have moderate room-temperature $Q$-factors below $2\cdot10^4$ (see \cref{App_g0}), which is much smaller than other optical cavities used for cavity optomechanics, including those that incorporate localized modes \cite{Eichenfield2009,Mitchell2014,Ghorbel2019,Arregui2021arXiv,Topolancik2007}. However, the strong field confinement in the slot leads to a considerable optomechanical coupling to the GHz acoustic modes, enabling their use as transducers of mechanical motion.\

\section{Mechanical characterization }
\label{OM_transduction}

We characterize the mechanical spectrum of the cavity-waveguides described above by using two different techniques. First, we use Brillouin Light Scattering (BLS) spectroscopy, where non-resonant light is focused onto the sample surface and non-linearly scattered by thermally activated acoustic phonons, which is then measured.\ The back-scattered signal from these acoustic phonons is collected and the vibrations manifest as peaks in the BLS spectrum. We previously reported on the use of this technique for mechanical dispersion mapping \cite{Florez2022}, which we apply here to obtain the mechanical band gap for each domain (shield and center), as well as the waveguide dispersion in the central region, as detailed in \Cref{PnC_shield}.\ The strength of this technique lies in its relative ease of incorporation as it is non-invasive and contactless. Further, it enables the detection of high frequency GHz guided mechanical modes at room temperature, without the need to drive an optical mode to observe mechanics and without the need for incorporating other on-chip architectures that may add complexity \cite{Florez2022,Fang2016,Patel2018,Liu2019}.\ \Cref{Fig3}a shows the BLS spectrum of the central region of the waveguide, measured using a 532 nm laser at an incident angle of 20.1$^\circ$, which corresponds to a wavevector of $k_x=0.7\pi$/$\ac$. The spectrum exhibits a broad ($\approx160$ MHz) peak at $\sim6.85$ GHz over a relatively flat band gap region spanning from 4.8 to 7.5 GHz. Comparing the BLS spectrum to the simulated band (\Cref{PnC_shield}) shows quantitative agreement between the probed structure and the simulated one, indicating that the measured mode at 6.8 GHz corresponds to the waveguide mode. This is also confirmed by focusing the laser beam within the crystal away from the waveguide, where the peak corresponding to the guided mode then vanishes. In principle, the BLS spectrum should reveal spectral features for all Fabry-Pérot acoustic modes having reciprocal space components including the $k$-value determined by the excitation angle. This should lead to a set of peaks over a bandwidth that decreases with waveguide length. However, only a single broad peak is observed. This is due to the limited spectral resolution of the BLS measurement ($\sim 100$ MHz) as well as to a relaxation of the phase-matching condition resulting from focusing of the laser light, which increases the probed wave-vectors to the full solid-angle of the incident beam. The measured spectrum is therefore an integration of all of the Brillouin scattering events occurring over a range $\Delta k_x$ around the targeted $k_x$.\\

\begin{figure}[!ht]
\centering
\includegraphics[trim={0.05cm 0.5cm 0cm 0.2cm},clip]{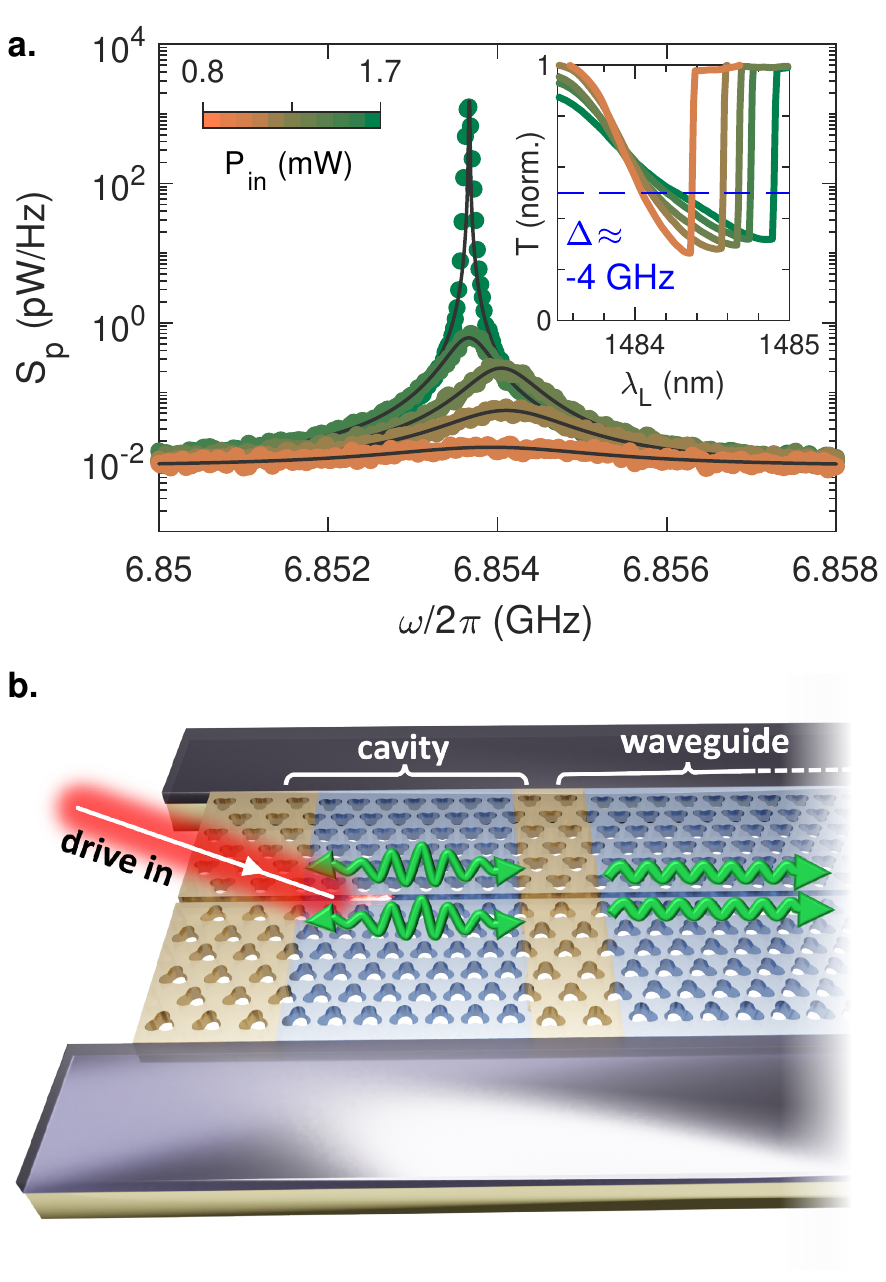}
\caption{\textbf{Optomechanical amplification of a GHz mechanical mode.} (a) Radio-frequency spectrum of the transmitted light showing the mechanical mode.\ Dynamical back-action is observed, which leads to mechanical lasing.\ The color scale indicates the laser power in mW.\ The inset shows the driven optical mode which exhibits a greater thermo-optic shift as the power is increased. (b) Schematic illustration of a "phononic laser", where acoustic waves (green arrows) are generated in the cavity which is side-coupled to a waveguide.}
\label{Fig4}
\end{figure}

The second technique we employ for mechanical characterization is the measurement of the optomechanically-induced transmittance fluctuations of a monochromatic light source driving the localized optical cavities.  \Cref{Fig3}b shows the RF spectrum detected with an electronic spectrum analyzer (ESA) when a localized mode with central wavelength $\lambda_L=1484.5$ nm is driven with a blue-detuned laser. The signal-to-noise ratio (SNR) is calculated from the power spectral density by normalizing to the noise level measured when the laser is off. To allow for a direct comparison with the BLS measurement in \Cref{Fig3}a, the spectrum is shown over the same frequency range. The inset in \Cref{Fig3}b shows a zoom-in around 6.76 GHz which, unlike the BLS spectrum, shows overlapping spectral lines corresponding to different mechanical modes supported by the waveguide over a bandwidth of $\approx 110$ MHz.\ These are roughly equi-spaced in frequency, and possibly correspond to acoustic Fabry-Pérot modes (such as those shown in simulations in \Cref{Fig1}c). However, the bandwidth over which they are observed is much smaller than the full frequency span of the guided mode, indicating that the optomechanical couplings $g_0$ to the driven optical mode are too low in many cases. This is further supported by the relative intensities among the transduced peaks and through the data shown in \Cref{Fig3}c and d, which respectively show the optical transmission spectrum and the associated mechanical RF transduction spectra when four of the localized modes are driven (as given by the dashed lines in \Cref{Fig3}c). The localized nature of the optical modes is revealed in several ways. First, a strong thermo-optic nonlinearity results in a significant red-shift and in sawtooth-shaped lineshapes in \Cref{Fig3}c. It also manifests in the strength of the optomechanically-transduced mechanical modes. For optical modes below 1484 nm, in the Fabry-Pérot regime, no mechanical modes are observed above the noise floor. The transduction intensity generally increases when the laser drives an optical mode at longer wavelengths. This results from a combination of the reduced mode volumes, an increase in quality factor of the localized optical modes closer to the cut-off of the band, and a stronger optomechanical coupling strength \cite{scullion2015enhancement}. In addition, a comparison of the spectra in \Cref{Fig3}d indicates that part of the observed mechanical peaks span more than one spectrum but may not always appear due to a different transduction strength which in turn depends on the driven optical mode.\ Similar results were obtained for cavities of varying lengths (see \cref{cav_length}).\\

The strong confinement of the optical field in the air-slot and of the mechanical displacement to the waveguide region, enables dynamical back-action and subsequent mechanical lasing, indicated by the shifting, narrowing, and increase in amplitude of the mechanical peak with increasing laser power (\Cref{Fig4}a). This occurs despite the relatively low values of the measured optical $Q$-factor.\ The color scale in \Cref{Fig4}a indicates the input laser power and the inset shows the transmittance across the optical resonance as the laser power is varied, with high powers leading to a greater thermo-optic shift. The transduced spectra for each power are acquired at a fixed coupling fraction, approximately corresponding to a detuning $\Delta = -4$ GHz. Even at relatively low input optical powers of $\sim$1.6 mW, high amplitude mechanical lasing in the GHz is observed.\ In fact, depending on the driven optical mode, different Fabry-Pérot acoustic modes (\Cref{Fig3}d) can be made to lase.
For the particular optical and mechanical mode of \Cref{Fig4}a, we extract an optomechanical coupling strength, $g_0/2\pi$, of 135 kHz from a linear fit to the evolution of the mechanical linewidth with input optical power \cite{Aspelmeyer2014} (see \cref{App_g0}).\ This assumes that optomechanical damping is the only source of linewidth narrowing, a hypothesis that is confirmed by calibration of the $g_0/2\pi$ using an alternative method relying on the use of an electro-optic phase modulator ($\phi$-EOM) (see \cref{phase_eom}), the results of which further confirm the localized nature of the optical modes in this platform.\ 


\section{Conclusion}


In summary, we present a 2D suspended silicon phononic crystal waveguide sustaining a single ($z$-symmetric) acoustic band near 7 GHz that sits within a 2.3 GHz-wide mechanical band gap, which we confirm using BLS spectroscopy. To enable the generation of coherent acoustic phonons propagating along this waveguide mode, we transform the structure into an efficient optomechanical waveguide by mirroring the PnC and leaving an air-slot in the middle between them. This results in a slot photonic-crystal waveguide supporting a slot-guided optical mode at telecom wavelengths. We then exploit localized optical modes spontaneously formed due to sidewall roughness to couple to the original phononic waveguides on both sides of the slot. The optomechanical coupling strengths achieved are enough to enable coherent phonon generation at relatively low input powers directly within the waveguide. This is beneficial compared to previously employed optomechanical generation schemes that use a carefully engineered optomechanical cavity at the edge of the phononic waveguide of interest. This concept is schematically illustrated in \Cref{Fig4}b for the structure presented here. In this case, the generation bandwidth is not limited to the linewidth of a single mechanical cavity mode but to multiple mechanical modes which can be selectively driven using specific localized optical modes and then transferred into a perfectly mode-matched semi-infinite waveguide.\

A peculiarity of the design presented in this work stems from the utilization of two mechanical waveguides separated by an air-slot. In this case, the system can be visualized as two individual plates that are separated by an air gap. The separation of these plates means that they are mechanically-uncoupled, but sustain mechanical modes that both couple to the same localized optical mode. Such a mechanical-optical-mechanical modal configuration has previously been explored in a number of platforms \cite{Zhang2012,Dong:14,ShkarinPRL2014} including zipper nanobeam structures that also exploit field enhancement effects in air slots \cite{Grutter:15}, although the design here extends this deeper into the GHz regime. 
The physical properties that are specifically encountered in multimode optomechanical systems lead to a wide variety of interesting phenomena such as synchronization \cite{Colombano2019}, exceptional points \cite{PREDjorwe,delPino2022}, or non-reciprocital states \cite{ruesink2018optical,de2020nonreciprocal}, which each may serve to enrich the breadth of integrated phononics.


\begin{acknowledgments}
We acknowledge the support from the project LEIT funded by the European Research Council, H2020 Grant Agreement No. 885689. ICN2 is supported by the Severo Ochoa program from the Spanish Research Agency (AEI, grant no. SEV-2017-0706) and by the CERCA Programme / Generalitat de Catalunya. R.C.N. acknowledges funding from the EU-H2020 research and innovation programme under the Marie Sklodowska Curie Individual Fellowship (Grant No. 897148). O.F. is supported by a BIST PhD fellowship under the Marie Sklodowska Curie grant agreement (No. 754558). M.A. and S.S. gratefully acknowledge funding from the Villum Foundation Young Investigator Program (Grant No.\ 13170), the Danish National Research Foundation (Grant No.\ DNRF147 - NanoPhoton), Innovation Fund Denmark (Grant No.\ 0175-00022 - NEXUS), and Independent Research Fund Denmark (Grant No.\ 0135-00315 - VAFL).
\end{acknowledgments}

\appendix

\section{Photonic design}
\label{SI:Photonics}

\begin{figure}[!ht]
\centering
\includegraphics[trim={0.2cm 0.3cm 0.4cm 0.4cm},clip]{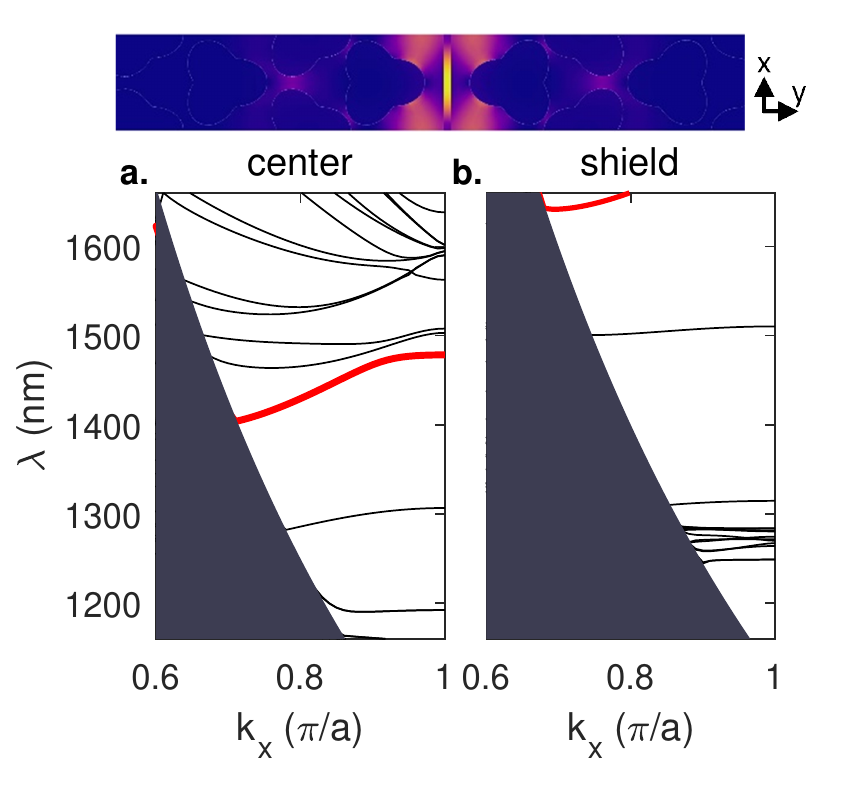}
\caption{\textbf{Photonic crystal design}. Photonic band structures of (a) the center region (left) and (b) the shield region (right), with the slot-confined guided mode highlighted in red, and the light cones shaded in dark grey.\ The simulated electromagnetic energy of the guided mode, which shows that it is confined inside the slot, is shown above the band structures.}
\label{SI_PhC}
\end{figure}

In addition to confining the mechanical modes to the length of the waveguide, the shield region also confines the optical modes.\ The optical band structures for the center and shield regions are calculated with FEM simulations (\Cref{SI_PhC}a and \Cref{SI_PhC}b, respectively).\ The guided mode within each region is highlighted in red.\ The light cone is shaded in dark grey. The slot-confined optical guided mode is not perfectly decoupled from the bulk bands, which results in the relatively poor optical Q-factors for the Fabry-Pérot optical modes. However, the guided modes found in the two regions do not overlap in frequency, and thus they cannot couple to one another.\ The associated mode profile for this mode for the center region is shown in \Cref{SI_PhC} (top), showing that the field is confined in the slot.\ Note that the axes are rotated by 90$^{o}$ for this mode profile.

\section{Fabrication}
\label{fabrication}

The sample is fabricated in chips cleaved from a commercial 12-inch silicon-on-insulator (SOI) wafer, which has a nominal 220 nm thick device layer and a 2 $\mu$m buried oxide.\ The lithography and release etch of the suspended membrane is carried out as detailed in \cite{Albrechtsen2021arXiv}.\ To reduce the sidewall roughness, we improve the mask selectivity by using a multi-layered hard mask, which we etch sequentially in the same reactor using modified versions of the CORE-process \cite{Nguyen2020,Albrechtsen2021arXiv,Nguyen2021}.\ In this process, oxygen is used to passivate the etch in the oxidation-step (O-step), the passivation at the bottom of all features is removed in a low-pressure SF$_6$ plasma in the remove-step (R-step), and the open features are etched isotopically in the etch-step (E-step).\

We first deposit 30 nm of poly-crystalline chromium followed by 30 nm of poly-crystalline silicon on the SOI chip without breaking vacuum to avoid oxidation of the chromium layer.\ We spin-coat $\sim50$ nm of chemically semi-amplified resist (AR-P CSAR6200.09 diluted 1:1 in anisole) and define the patterns with a 100 keV 100 MHz JEOL-9500FSZ electron-beam writer with a current of 195 pA and a dose density of 3 aC/nm$^2$ in the center.\ The pattern is etched into the poly-crystalline silicon hard mask from the CSAR soft mask using 8~cycles of a modified CORE process \cite{Albrechtsen2021arXiv}.\ In this modified process, the time of the E-step is reduced from 73 s $\rightarrow$ 20 s to reduce sidewall roughness and the passivation is reduced accordingly by decreasing the time of the O-step from 3 s $\rightarrow$ 2 s.\ Following this, the soft mask is removed in a 30 W oxygen plasma at 50 mTorr for 5 min.\ The gas flows and pressure here are the same as the O-step in the CORE-process.\ The pattern is then etched into the poly-crystalline chromium hard mask from the poly-crystalline silicon hard mask in 65 cycles of the unmodified chromium-etching CORE-process \cite{Nguyen2021}, which is an oxygen-rich plasma at 50 mTorr and 40 W, with 3 sccm of SF$_6$ flow for 1 s every 10 s.\
Next, the crystalline silicon device-layer (100) is etched from the chromium hard mask in 25 cycles of another modified CORE-process.\ For this modified process, the time of the E-step is reduced from 73 s $\rightarrow$ 20 s to reduce sidewall roughness similar to in the poly-crystalline silicon etch.\ Since this silicon layer is etched from a hard mask rather than a soft mask, this is compensated by increasing the platen power of the R-step from 10 W $\rightarrow$ 18 W, while keeping the passivation O-step at 3 s.\ This improves yield but was not possible when etching from a soft mask as the oxygen plasma of the O-step slowly etches the resist laterally (mask retraction) causing a change in feature size and therefore should be reduced as much as possible.\ Finally, the remaining chromium mask is stripped by submerging the chip for 10 min in commercial solution, Chrome Etch 18, which consists of nitric acid (HNO$_3$), perchloric acid (HClO$_4$), and ceric salts ((NH$_4$)$_2$Ce(NO$_3$)$_6$), followed by a 2 min dip in a Piranha solution (4:1 mix of H$_2$SO$_4$:H$_2$O$_2$).\ The structures are released by a 3.8 $\upmu$m isotropic underetch of the buried oxide in anhydrous hydrofluoric acid (99.995 \%) over 25 min in a temperature- and pressure-controlled process \cite{Albrechtsen2021arXiv} using an SPTS uEtch vapor process.\

\section{Determination of \texorpdfstring{$g_0$}{} by fitting the mechanical linewidth}
\label{App_g0}

\begin{figure}[!ht]
 \centering
  \includegraphics{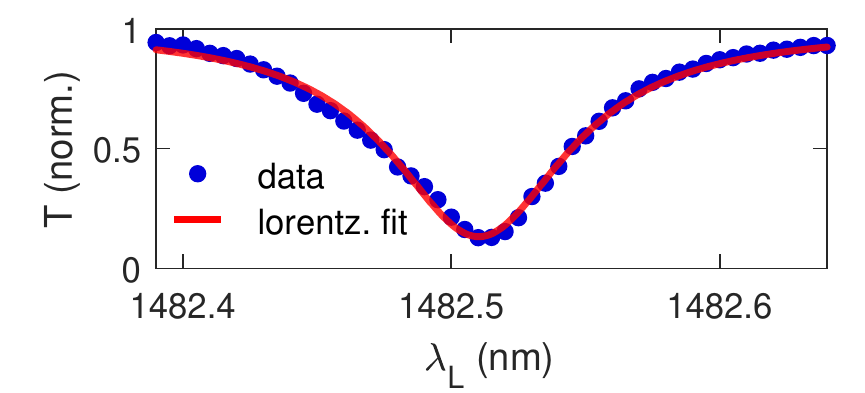}
\caption{\textbf{Optical Q-factor determination}. Localized optical mode transmission response (blue dots) measured with input power $<0.2$ mW. The dip is fitted with a Lorentzian function (red line) enabling the extraction of $\lambda_c=1482.510$ nm, $\kappa_i/2\pi=4.0$ GHz, and $\kappa_e/2\pi=7.0$ GHz. }
\label{SI_lin_OM}
\end{figure}

We also estimate the optomechanical coupling strength, $g_0$, by fitting the mechanical linewidth \cite{Arregui2021arXiv,Aspelmeyer2014}.\ The effective mechanical linewidth is given by:

\begin{equation}
\begin{split}
    \Gamma_\mathrm{eff} = \Gm + g_0^2|\overline{a}|^2 &\Big(\frac{\kappa_t}{(\Delta+\Wm)^2 + 
    \kappa_t^2/4} \\
    &- \frac{\kappa_t}{(\Delta-\Wm)^2 + \kappa_t^2/4}\Big)
    \end{split}
\end{equation}

where $\Wm/2\pi$ and $\Gm/2\pi$ are the mechanical resonance frequency and natural linewidth, respectively. $\Delta=\omega_L - \omega_c$ is the optical detuning, $\kappa_t$ is the cavity total optical linewidth, and $|\overline{a}|^2$ is the intracavity photon number given by

\begin{equation*}
\label{Geff_eq}
    |\overline{a}|^2 = \frac{\kappa_e}{\Delta^2 + \kappa_t^2/4}\frac{\pin}{\hbar\omega_L}
\end{equation*}

where $\kappa_e$ represents the cavity extrinsic loss rate. From this equation, the mechanical linewidth $\Gamma_\mathrm{eff}$ is directly proportional to the incident laser power $\pin$.\ The optomechanical coupling $g_0$ can be extracted from this fit by knowledge of all other parameters provided that they are kept constant.

We first calibrate the optical parameters by fitting the transmission response with the transmission function $T(\Delta)=(\Delta^2+\kappa_i^2)/(\Delta^2+\kappa_t^2)$ where the internal decay rate $\kappa_i$, the external decay rate $\kappa_e$, and the cavity resonance frequency $\omega_c$ are used as fitting parameters. The total decay rate of the cavity is: $\kappa_t=\kappa_i+\kappa_e$. The measurement (\Cref{SI_lin_OM}) is performed with input laser power $\pin<0.2$ mW, which is sufficiently low to remove the thermo-optical nonlinearities. We obtain $\omega_c/2\pi=202.36$ THz (i.e. $\lambda_c = 1482.510$ nm), $\kappa_i/2\pi=4.0$ GHz, and $\kappa_e/2\pi=7.0$ GHz. We deduce the localized optical mode loaded Q-factor, $Q_\mathrm{opt}=18,400$.

\begin{figure}[!ht]
 \centering
  \includegraphics{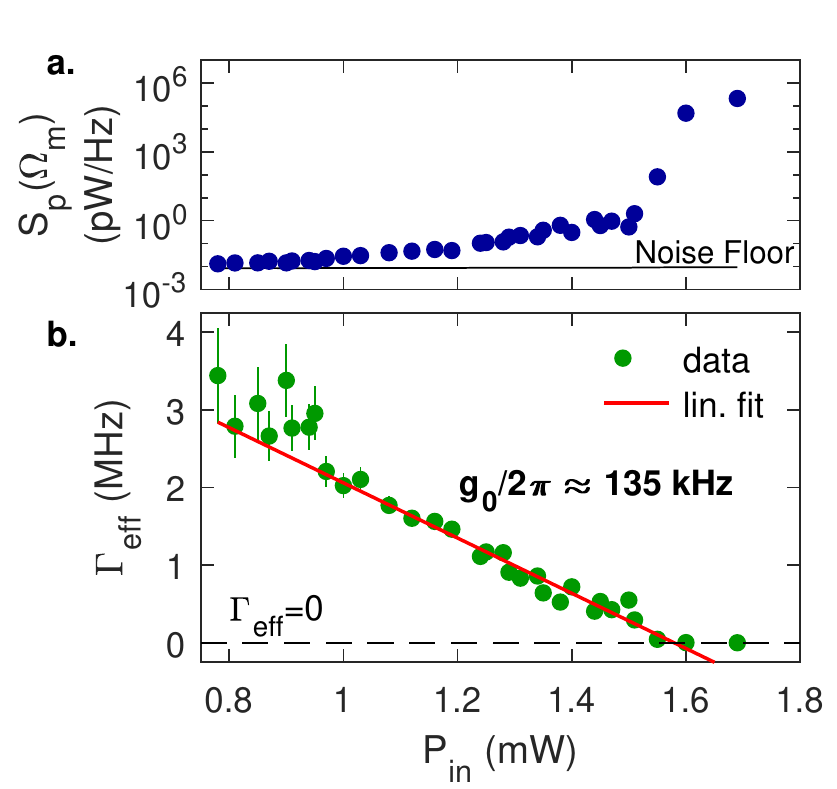}
\caption{\textbf{Vacuum optomechanical coupling calibration}. a) Mechanical peak power (blue dots) plotted with the detection noise floor (black line). b) Effective mechanical linewidth plotted as a function of the input laser power. The data is extracted from \Cref{Fig3}a. An optomechanical coupling rate $g_0/2\pi=135$ kHz is obtained from the linear fit of the mechanical linewidth (red line).}
\label{SI_g0}
\end{figure}

The same measurement is reproduced for increasing input power, as shown in the inset of \Cref{Fig4}a. Higher input power leads to a significant red-shift of the cavity resonance, which exhibits hysteretic behavior typical of the thermo-optical nonlinearities encountered in silicon photonic devices \cite{Navarro-Urrios2014,Maire2018}. We assume that \Cref{Geff_eq} remains valid despite this deviation from the linear optical cavity scenario and lock the detuning parameter by always setting the laser wavelength such that $T(\Delta)=0.5$, which corresponds to $\Delta\approx-4$ GHz in the linear regime. The RF transmission spectrum is recorded as shown in \Cref{Fig4}a. The mechanical resonance is fitted with a Lorentzian lineshape such that the mechanical peak amplitude $S_p(\Wm)$ and linewidth $\Gamma_\mathrm{eff}$ can be plotted as a function of the input laser power (see \Cref{SI_g0}a and \Cref{SI_g0}b, respectively). The detection noise floor is also shown in \Cref{SI_g0}a for reference. An exponential amplification of the mechanical mode is first observed, followed by a threshold around 1.58 mW. Above this threshold, the mode lies in the lasing regime. The mechanical linewidth is fitted below this threshold with a red line in \Cref{SI_g0}b. We deduce the vacuum optomechanical coupling rate $g_0/2\pi = 135$ kHz from the slope and the natural mechanical linewidth $\Gm/2\pi=5.6$ MHz.

\section{Phononic crystal characterization with Brillouin light scattering spectroscopy.}
\label{PnC_shield}

\begin{figure}[!ht]
 \centering
  \includegraphics[trim={0.25cm 0 0.15cm 0.1cm},clip]{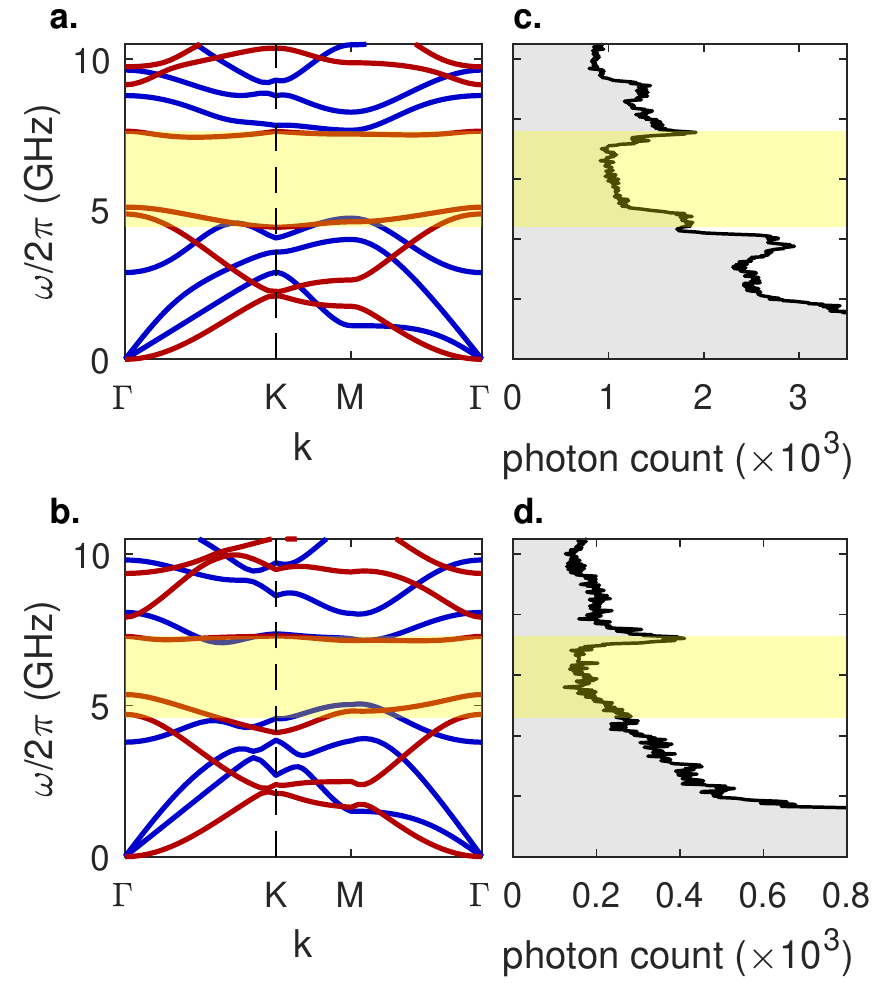}
\caption{\textbf{PnC characterization}. FEM simulation of the mechanical band diagram of the center (a) and the shield (b) domains. Symmetrical (antisymmetrical) modes are shown in blue (red). The band gap at the K point is highlighted in yellow. Experimental characterization of the center (c) and shield (d) domains using BLS spectroscopy performed at the K point.}
\label{SI_PnCshield}
\end{figure}

We compute the band diagram for a PnC for both the center and the shield domains using FEM simulations, shown in \Cref{SI_PnCshield}a and \Cref{SI_PnCshield}b, respectively. These simulations do not consider a waveguide, i.e., there is no guided mode. In both cases, we observe a full mechanical band gap. The band gap of the center region is centered at 6.2 GHz with a width of 2.3 GHz while that of the shield region is centered at 6.1 GHz with a width of 1.5 GHz. The band gap is highlighted in yellow.

We are able to characterize the thermal phonons in both domains by BLS spectroscopy. The BLS laser is focused in the center domain, away from the phononic waveguide. The measurement is performed at a specific angle, $\alpha=20.1^\circ$, which corresponds to the wavevector lying at the K point of the Brillouin zone (see dashed line in \Cref{SI_PnCshield}a and b). The data shown in \Cref{SI_PnCshield}a was acquired in approximately 12 hours (91,000 photon counts). The same measurement is performed by focusing the laser in the shield region (see \Cref{SI_PnCshield}b) with about 29,000 photon counts.\ This long acquisition time is required due to the relatively weak Brillouin scattering signal compared to elastic Rayleigh scattering.\ The band gap at the K point (highlighted in yellow) in \Cref{SI_PnCshield}a and b are also placed over the same frequency range as their associated BLS measurements in \Cref{SI_PnCshield}c and d and show good agreement for both of the regions between simulation and measurement.

\begin{figure}[!ht]
\centering
\includegraphics[trim={0cm 0.2cm 0cm 0cm},clip]{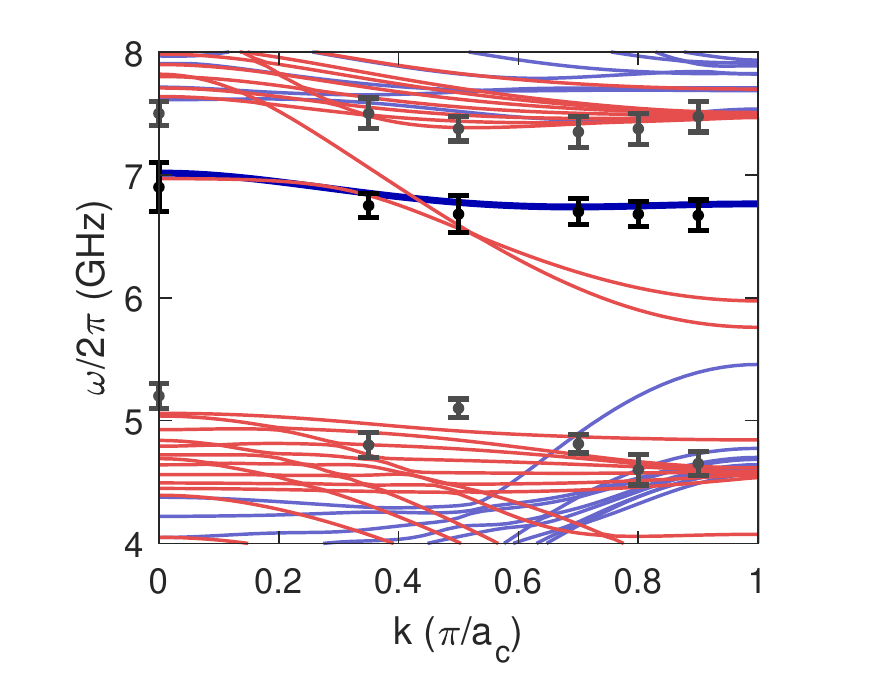}
\caption{\textbf{BLS waveguide characterization.} Experimental measurements of the guided mode (dark dots) and of the band gap edges (dark grey dots) compared with the simulated symmetrical  and antisymmetrical mechanical modes, shown with blue and red lines, respectively. Error bars indicate the BLS measured peak linewidths.}
\label{SI_simu_BLS}
\end{figure}

When the laser spot is placed at the center of the waveguide within the center region of the crystal, the thermally excited mechanical mode that was identified as the guided acoustic mode can be detected, as shown in \cref{Fig3}a. Different measurements are taken at this position for various wavevectors by changing the incident angle. For example, an incident angle of 20.1$^\circ$ corresponds to a wavevector of $k_x = 0.7\pi/a$. In doing so, the dispersion relation can be experimentally mapped (\Cref{SI_simu_BLS}, and allow for a direct comparison of the experimental measurements (points) with the FEM simulations (lines). The grey and black points correspond to the band gap edges and the guided mode, respectively. A close quantitative and qualitative correspondence between the observed phononic elastic dispersion relation and the associated simulation is observed.

\section{Cavity length}
\label{cav_length}

\begin{figure*}[!ht]
 \centering
  \includegraphics[trim={0.2cm 0 0.2cm 0cm},clip]{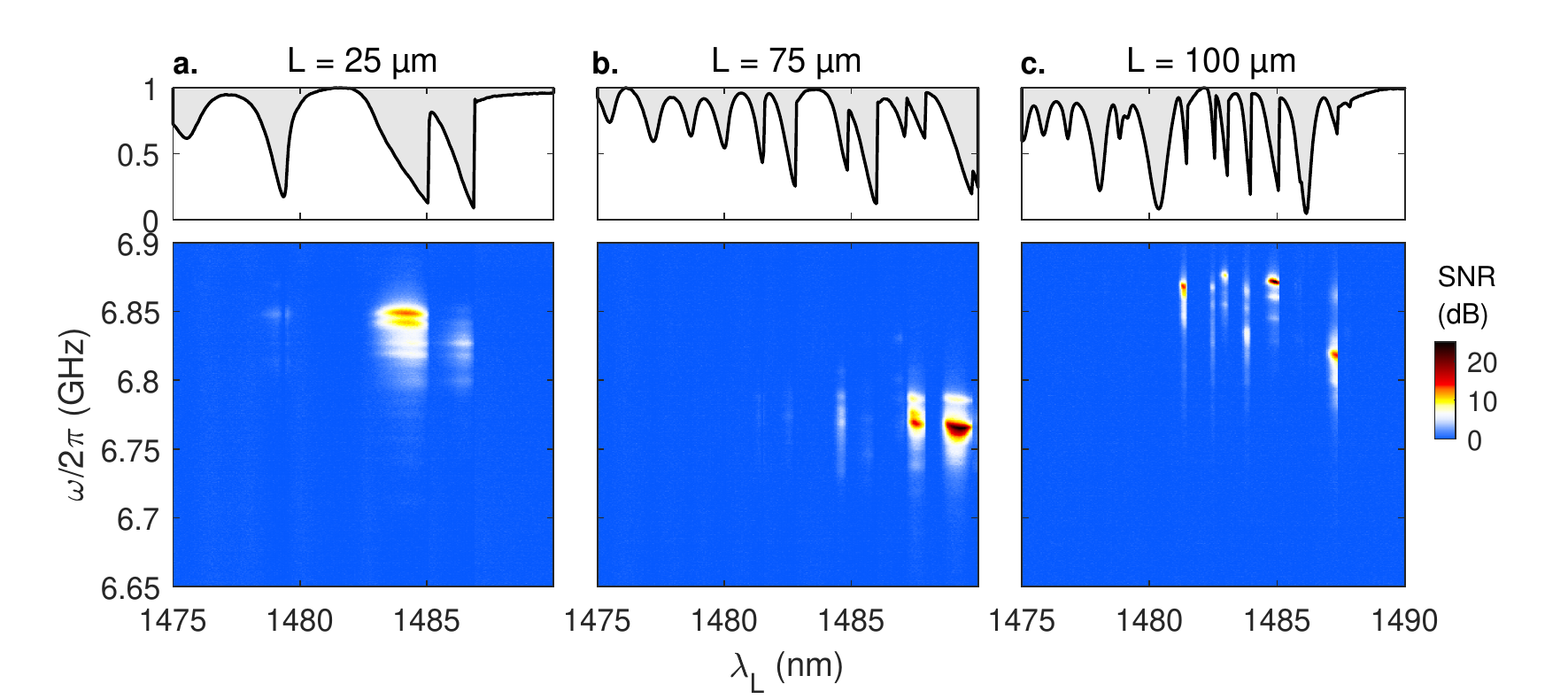}
\caption{\textbf{Cavity length dependence of optical and mechanical spectra}. The optical transmission spectrum (top) and associated mechanical spectrum (bottom) for three different cavity lengths of a) L=25 $\um$, b) L=75 $\um$, and c) L=100 $\um$.}
\label{SI_length}
\end{figure*}

The cavity length plays a crucial role in the spectral distributions of both photonic and phononic modes.\ 
We measure the optical transmission (\Cref{SI_length}a) for three different structures of varying length ($L$ = 25 $\mu$m, $L$ = 75 $\mu$m, and $L$ = 100 $\mu$m), but with the same geometric parameters as specified in \Cref{design}.\ There is a tendency towards a greater optical density of states and a narrowing of the modes with increasing cavity length. Each of the spectra are taken with a different input optical power of $\pin=5$ mW, $\pin=4$ mW, and $\pin=1.4$ mW from shortest to longest cavity length.\ For each optical transmission spectrum, the mechanical noise spectrum is also measured as a function of the laser wavelength.\ The spectra are normalized by the frequency-dependent noise floor of the ESA, leading to the SNR as indicated by the color scale (\Cref{SI_length}b).\ Mechanical modes in the phononic band gap are transduced in the laser field when an optical mode is driven.\ This transduction increases significantly when driving localized optical modes, i.e. where the optical band flattens at the Brillouin zone edge, as discussed in \Cref{OM_transduction}.\
Both the optical and mechanical modes of the $L$ = 75 $\mu$m cavity are spectrally shifted, with red-shifted optical modes and mechanical modes shifted to lower frequency, relative to the spectra of the other two cavity lengths.\ This shift has gradually increased over the time scale of months over which the sample has undergone measurement, and likely results from slight modification of the structure due to long illumination times under high powers in BLS spectroscopy.

\section{Determination of \texorpdfstring{$g_0$}{} with a phase electro-optic modulator}
\label{phase_eom}

To further study the localized nature of the modes in this system, the $g_0/2\pi$ was measured at different positions along the waveguide by calibration with a phase electro-optic modulator (EOM) (Table \ref{table:1}). The calibration of the optomechanical coupling using a phase EOM \cite{Gorodetsky2010} is made in the absence of dynamical back-action.\ The four different positions in Table \ref{table:1} correspond with the three positions indicated in \Cref{Fig3}b, along with an additional one at $\pos=12$ $\um$.\ At each position, a localized optically driven mode with wavelength, $\lambda_L$, excites the mechanical mode at frequency, $\Wm$, with a mechanical dissipation rate, $\Gm$. The different optomechanical couplings (i.e. varying $g_0/2\pi$) at various positions along the waveguide at slightly different values of $\lambda_L$ but roughly the same value of $\Wm$ further suggests localized behavior (notably, excitation of a mechanical mode driven by a localized optical mode).\ Furthermore, these results confirm the previous calibration of $g_0$ (appendix \ref{App_g0}).\

\begin{table}[!ht]
\centering
\begin{tabular}{c c c c c} 
 \hline\hline
 $\pos$ ($\um$) & $g_0/2\pi$ (kHz) & $\lambda_{\textrm{L}}$ (nm) & $\Omega_{\textrm{m}}$ (GHz) & $\Gamma_{\textrm{m}}$ (MHz) \\ [0.5ex] 
 \hline
 12 & 72 $\pm$ 4 & 1484.14 & 6.845 & 4.8 \\
 19 & 104 $\pm$ 6 & 1483.47 & 6.849 & 10 \\
 35 & 169 $\pm$ 10 & 1484.05 & 6.846 & 2.5 \\
 60 & 124 $\pm$ 8 & 1485.85 & 6.846 & 13.4 \\ [1ex] 
 \hline\hline
\end{tabular}
\caption{Phase electro-optic modulator measured $g_0/2\pi$ at the four different positions along the waveguide}
\label{table:1}
\end{table}

\clearpage
\bibliography{biblio}

\begin{thebibliography}{60}%
\makeatletter
\providecommand \@ifxundefined [1]{%
 \@ifx{#1\undefined}
}%
\providecommand \@ifnum [1]{%
 \ifnum #1\expandafter \@firstoftwo
 \else \expandafter \@secondoftwo
 \fi
}%
\providecommand \@ifx [1]{%
 \ifx #1\expandafter \@firstoftwo
 \else \expandafter \@secondoftwo
 \fi
}%
\providecommand \natexlab [1]{#1}%
\providecommand \enquote  [1]{``#1''}%
\providecommand \bibnamefont  [1]{#1}%
\providecommand \bibfnamefont [1]{#1}%
\providecommand \citenamefont [1]{#1}%
\providecommand \href@noop [0]{\@secondoftwo}%
\providecommand \href [0]{\begingroup \@sanitize@url \@href}%
\providecommand \@href[1]{\@@startlink{#1}\@@href}%
\providecommand \@@href[1]{\endgroup#1\@@endlink}%
\providecommand \@sanitize@url [0]{\catcode `\\12\catcode `\$12\catcode
  `\&12\catcode `\#12\catcode `\^12\catcode `\_12\catcode `\%12\relax}%
\providecommand \@@startlink[1]{}%
\providecommand \@@endlink[0]{}%
\providecommand \url  [0]{\begingroup\@sanitize@url \@url }%
\providecommand \@url [1]{\endgroup\@href {#1}{\urlprefix }}%
\providecommand \urlprefix  [0]{URL }%
\providecommand \Eprint [0]{\href }%
\providecommand \doibase [0]{http://dx.doi.org/}%
\providecommand \selectlanguage [0]{\@gobble}%
\providecommand \bibinfo  [0]{\@secondoftwo}%
\providecommand \bibfield  [0]{\@secondoftwo}%
\providecommand \translation [1]{[#1]}%
\providecommand \BibitemOpen [0]{}%
\providecommand \bibitemStop [0]{}%
\providecommand \bibitemNoStop [0]{.\EOS\space}%
\providecommand \EOS [0]{\spacefactor3000\relax}%
\providecommand \BibitemShut  [1]{\csname bibitem#1\endcsname}%
\let\auto@bib@innerbib\@empty
\bibitem [{\citenamefont {Wang}\ and\ \citenamefont {Li}(2007)}]{Wang2007}%
  \BibitemOpen
  \bibfield  {author} {\bibinfo {author} {\bibfnamefont {L.}~\bibnamefont
  {Wang}}\ and\ \bibinfo {author} {\bibfnamefont {B.}~\bibnamefont {Li}},\
  }\href {\doibase 10.1103/PhysRevLett.99.177208} {\bibfield  {journal}
  {\bibinfo  {journal} {Phys. Rev. Lett.}\ }\textbf {\bibinfo {volume} {99}},\
  \bibinfo {pages} {177208} (\bibinfo {year} {2007})}\BibitemShut {NoStop}%
\bibitem [{\citenamefont {Li}\ \emph {et~al.}(2012)\citenamefont {Li},
  \citenamefont {Ren}, \citenamefont {Wang}, \citenamefont {Zhang},
  \citenamefont {H{\"{a}}nggi},\ and\ \citenamefont {Li}}]{Li2012}%
  \BibitemOpen
  \bibfield  {author} {\bibinfo {author} {\bibfnamefont {N.}~\bibnamefont
  {Li}}, \bibinfo {author} {\bibfnamefont {J.}~\bibnamefont {Ren}}, \bibinfo
  {author} {\bibfnamefont {L.}~\bibnamefont {Wang}}, \bibinfo {author}
  {\bibfnamefont {G.}~\bibnamefont {Zhang}}, \bibinfo {author} {\bibfnamefont
  {P.}~\bibnamefont {H{\"{a}}nggi}}, \ and\ \bibinfo {author} {\bibfnamefont
  {B.}~\bibnamefont {Li}},\ }\href {\doibase 10.1103/RevModPhys.84.1045}
  {\bibfield  {journal} {\bibinfo  {journal} {Rev. Mod. Phys.}\ }\textbf
  {\bibinfo {volume} {84}},\ \bibinfo {pages} {1045} (\bibinfo {year}
  {2012})}\BibitemShut {NoStop}%
\bibitem [{\citenamefont {Sklan}(2015)}]{Sklan2015}%
  \BibitemOpen
  \bibfield  {author} {\bibinfo {author} {\bibfnamefont {S.~R.}\ \bibnamefont
  {Sklan}},\ }\href {\doibase 10.1063/1.4919584} {\bibfield  {journal}
  {\bibinfo  {journal} {AIP Advances}\ }\textbf {\bibinfo {volume} {5}},\
  \bibinfo {pages} {053302} (\bibinfo {year} {2015})},\ \Eprint
  {http://arxiv.org/abs/https://doi.org/10.1063/1.4919584}
  {https://doi.org/10.1063/1.4919584} \BibitemShut {NoStop}%
\bibitem [{\citenamefont {Ng}\ \emph {et~al.}(2022)\citenamefont {Ng},
  \citenamefont {Sachat}, \citenamefont {Cespedes}, \citenamefont {Poblet},
  \citenamefont {Madiot}, \citenamefont {Jaramillo-Fernandez}, \citenamefont
  {Xiao}, \citenamefont {Florez}, \citenamefont {Sledzinska}, \citenamefont
  {Sotomayor-Torres},\ and\ \citenamefont {Chavez-Angel}}]{Ng2022arXiv}%
  \BibitemOpen
  \bibfield  {author} {\bibinfo {author} {\bibfnamefont {R.~C.}\ \bibnamefont
  {Ng}}, \bibinfo {author} {\bibfnamefont {A.~E.}\ \bibnamefont {Sachat}},
  \bibinfo {author} {\bibfnamefont {F.}~\bibnamefont {Cespedes}}, \bibinfo
  {author} {\bibfnamefont {M.}~\bibnamefont {Poblet}}, \bibinfo {author}
  {\bibfnamefont {G.}~\bibnamefont {Madiot}}, \bibinfo {author} {\bibfnamefont
  {J.}~\bibnamefont {Jaramillo-Fernandez}}, \bibinfo {author} {\bibfnamefont
  {P.}~\bibnamefont {Xiao}}, \bibinfo {author} {\bibfnamefont {O.}~\bibnamefont
  {Florez}}, \bibinfo {author} {\bibfnamefont {M.}~\bibnamefont {Sledzinska}},
  \bibinfo {author} {\bibfnamefont {C.}~\bibnamefont {Sotomayor-Torres}}, \
  and\ \bibinfo {author} {\bibfnamefont {E.}~\bibnamefont {Chavez-Angel}},\
  }\href@noop {} {\enquote {\bibinfo {title} {Excitation and detection of
  acoustic phonons in nanoscale systems},}\ } (\bibinfo {year} {2022}),\
  \Eprint {http://arxiv.org/abs/2203.15476} {arXiv:2203.15476
  [cond-mat.mtrl-sci]} \BibitemShut {NoStop}%
\bibitem [{\citenamefont {Zivari}\ \emph {et~al.}(2021)\citenamefont {Zivari},
  \citenamefont {Stockill}, \citenamefont {Fiaschi},\ and\ \citenamefont
  {Gr\"oblacher}}]{Zivari2021arXiv}%
  \BibitemOpen
  \bibfield  {author} {\bibinfo {author} {\bibfnamefont {A.}~\bibnamefont
  {Zivari}}, \bibinfo {author} {\bibfnamefont {R.}~\bibnamefont {Stockill}},
  \bibinfo {author} {\bibfnamefont {N.}~\bibnamefont {Fiaschi}}, \ and\
  \bibinfo {author} {\bibfnamefont {S.}~\bibnamefont {Gr\"oblacher}},\
  }\href@noop {} {\enquote {\bibinfo {title} {{Non-classical mechanical states
  guided in a phononic waveguide}},}\ } (\bibinfo {year} {2021}),\ \Eprint
  {http://arxiv.org/abs/2108.06248} {arXiv:2108.06248 [cond-mat.mes-hall]}
  \BibitemShut {NoStop}%
\bibitem [{\citenamefont {Zivari}\ \emph
  {et~al.}(2022{\natexlab{a}})\citenamefont {Zivari}, \citenamefont {Fiaschi},
  \citenamefont {Burgwal}, \citenamefont {Verhagen}, \citenamefont {Stockill},\
  and\ \citenamefont {Gr\"oblacher}}]{Zivari2022arXiv}%
  \BibitemOpen
  \bibfield  {author} {\bibinfo {author} {\bibfnamefont {A.}~\bibnamefont
  {Zivari}}, \bibinfo {author} {\bibfnamefont {N.}~\bibnamefont {Fiaschi}},
  \bibinfo {author} {\bibfnamefont {R.}~\bibnamefont {Burgwal}}, \bibinfo
  {author} {\bibfnamefont {E.}~\bibnamefont {Verhagen}}, \bibinfo {author}
  {\bibfnamefont {R.}~\bibnamefont {Stockill}}, \ and\ \bibinfo {author}
  {\bibfnamefont {S.}~\bibnamefont {Gr\"oblacher}},\ }\href@noop {} {\enquote
  {\bibinfo {title} {{On-chip distribution of quantum information using
  traveling phonons}},}\ } (\bibinfo {year} {2022}{\natexlab{a}}),\ \Eprint
  {http://arxiv.org/abs/2204.05066} {arXiv:2204.05066 [cond-mat.mes-hall]}
  \BibitemShut {NoStop}%
\bibitem [{\citenamefont {Hatanaka}\ \emph {et~al.}(2014)\citenamefont
  {Hatanaka}, \citenamefont {Mahboob}, \citenamefont {Onomitsu},\ and\
  \citenamefont {Yamaguchi}}]{Hatanaka2014}%
  \BibitemOpen
  \bibfield  {author} {\bibinfo {author} {\bibfnamefont {D.}~\bibnamefont
  {Hatanaka}}, \bibinfo {author} {\bibfnamefont {I.}~\bibnamefont {Mahboob}},
  \bibinfo {author} {\bibfnamefont {K.}~\bibnamefont {Onomitsu}}, \ and\
  \bibinfo {author} {\bibfnamefont {H.}~\bibnamefont {Yamaguchi}},\ }\href
  {\doibase 10.1038/nnano.2014.107} {\bibfield  {journal} {\bibinfo  {journal}
  {Nature nanotechnology}\ }\textbf {\bibinfo {volume} {9}},\ \bibinfo {pages}
  {520} (\bibinfo {year} {2014})}\BibitemShut {NoStop}%
\bibitem [{\citenamefont {{Van Laer}}\ \emph {et~al.}(2015)\citenamefont {{Van
  Laer}}, \citenamefont {Kuyken}, \citenamefont {{Van Thourhout}},\ and\
  \citenamefont {Baets}}]{Laer2015}%
  \BibitemOpen
  \bibfield  {author} {\bibinfo {author} {\bibfnamefont {R.}~\bibnamefont {{Van
  Laer}}}, \bibinfo {author} {\bibfnamefont {B.}~\bibnamefont {Kuyken}},
  \bibinfo {author} {\bibfnamefont {D.}~\bibnamefont {{Van Thourhout}}}, \ and\
  \bibinfo {author} {\bibfnamefont {R.}~\bibnamefont {Baets}},\ }\href
  {\doibase 10.1038/nphoton.2015.11} {\bibfield  {journal} {\bibinfo  {journal}
  {Nature Photonics}\ }\textbf {\bibinfo {volume} {9}},\ \bibinfo {pages} {199}
  (\bibinfo {year} {2015})}\BibitemShut {NoStop}%
\bibitem [{\citenamefont {Fu}\ \emph {et~al.}(2019)\citenamefont {Fu},
  \citenamefont {Shen}, \citenamefont {Xu}, \citenamefont {Zou}, \citenamefont
  {Cheng}, \citenamefont {Han},\ and\ \citenamefont {Tang}}]{Fu2019}%
  \BibitemOpen
  \bibfield  {author} {\bibinfo {author} {\bibfnamefont {W.}~\bibnamefont
  {Fu}}, \bibinfo {author} {\bibfnamefont {Z.}~\bibnamefont {Shen}}, \bibinfo
  {author} {\bibfnamefont {Y.}~\bibnamefont {Xu}}, \bibinfo {author}
  {\bibfnamefont {C.-L.}\ \bibnamefont {Zou}}, \bibinfo {author} {\bibfnamefont
  {R.}~\bibnamefont {Cheng}}, \bibinfo {author} {\bibfnamefont
  {X.}~\bibnamefont {Han}}, \ and\ \bibinfo {author} {\bibfnamefont {H.~X.}\
  \bibnamefont {Tang}},\ }\href {\doibase 10.1038/s41467-019-10852-3}
  {\bibfield  {journal} {\bibinfo  {journal} {Nature Communications}\ }\textbf
  {\bibinfo {volume} {10}},\ \bibinfo {pages} {2743} (\bibinfo {year}
  {2019})}\BibitemShut {NoStop}%
\bibitem [{\citenamefont {Su}\ \emph {et~al.}(2010)\citenamefont {Su},
  \citenamefont {Olsson~III}, \citenamefont {Leseman},\ and\ \citenamefont
  {El-Kady}}]{su2010realization}%
  \BibitemOpen
  \bibfield  {author} {\bibinfo {author} {\bibfnamefont {M.}~\bibnamefont
  {Su}}, \bibinfo {author} {\bibfnamefont {R.}~\bibnamefont {Olsson~III}},
  \bibinfo {author} {\bibfnamefont {Z.}~\bibnamefont {Leseman}}, \ and\
  \bibinfo {author} {\bibfnamefont {I.}~\bibnamefont {El-Kady}},\ }\href@noop
  {} {\bibfield  {journal} {\bibinfo  {journal} {Applied Physics Letters}\
  }\textbf {\bibinfo {volume} {96}},\ \bibinfo {pages} {053111} (\bibinfo
  {year} {2010})}\BibitemShut {NoStop}%
\bibitem [{\citenamefont {Soliman}\ \emph {et~al.}(2010)\citenamefont
  {Soliman}, \citenamefont {Su}, \citenamefont {Leseman}, \citenamefont
  {Reinke}, \citenamefont {El-Kady},\ and\ \citenamefont
  {Olsson~III}}]{soliman2010phononic}%
  \BibitemOpen
  \bibfield  {author} {\bibinfo {author} {\bibfnamefont {Y.}~\bibnamefont
  {Soliman}}, \bibinfo {author} {\bibfnamefont {M.}~\bibnamefont {Su}},
  \bibinfo {author} {\bibfnamefont {Z.}~\bibnamefont {Leseman}}, \bibinfo
  {author} {\bibfnamefont {C.}~\bibnamefont {Reinke}}, \bibinfo {author}
  {\bibfnamefont {I.}~\bibnamefont {El-Kady}}, \ and\ \bibinfo {author}
  {\bibfnamefont {R.}~\bibnamefont {Olsson~III}},\ }\href@noop {} {\bibfield
  {journal} {\bibinfo  {journal} {Applied physics letters}\ }\textbf {\bibinfo
  {volume} {97}},\ \bibinfo {pages} {193502} (\bibinfo {year}
  {2010})}\BibitemShut {NoStop}%
\bibitem [{\citenamefont {Pourabolghasem}\ \emph {et~al.}(2018)\citenamefont
  {Pourabolghasem}, \citenamefont {Dehghannasiri}, \citenamefont {Eftekhar},\
  and\ \citenamefont {Adibi}}]{Pourabolghasem2018}%
  \BibitemOpen
  \bibfield  {author} {\bibinfo {author} {\bibfnamefont {R.}~\bibnamefont
  {Pourabolghasem}}, \bibinfo {author} {\bibfnamefont {R.}~\bibnamefont
  {Dehghannasiri}}, \bibinfo {author} {\bibfnamefont {A.~A.}\ \bibnamefont
  {Eftekhar}}, \ and\ \bibinfo {author} {\bibfnamefont {A.}~\bibnamefont
  {Adibi}},\ }\href {\doibase 10.1103/PhysRevApplied.9.014013} {\bibfield
  {journal} {\bibinfo  {journal} {Phys. Rev. Applied}\ }\textbf {\bibinfo
  {volume} {9}},\ \bibinfo {pages} {014013} (\bibinfo {year}
  {2018})}\BibitemShut {NoStop}%
\bibitem [{\citenamefont {Dehghannasiri}\ \emph {et~al.}(2018)\citenamefont
  {Dehghannasiri}, \citenamefont {Eftekhar},\ and\ \citenamefont
  {Adibi}}]{Dehghannasiri2018}%
  \BibitemOpen
  \bibfield  {author} {\bibinfo {author} {\bibfnamefont {R.}~\bibnamefont
  {Dehghannasiri}}, \bibinfo {author} {\bibfnamefont {A.~A.}\ \bibnamefont
  {Eftekhar}}, \ and\ \bibinfo {author} {\bibfnamefont {A.}~\bibnamefont
  {Adibi}},\ }\href {\doibase 10.1103/PhysRevApplied.10.064019} {\bibfield
  {journal} {\bibinfo  {journal} {Phys. Rev. Applied}\ }\textbf {\bibinfo
  {volume} {10}},\ \bibinfo {pages} {064019} (\bibinfo {year}
  {2018})}\BibitemShut {NoStop}%
\bibitem [{\citenamefont {Florez}\ \emph {et~al.}(2022)\citenamefont {Florez},
  \citenamefont {Arregui}, \citenamefont {Albrechtsen}, \citenamefont {Ng},
  \citenamefont {Gomis-Bresco}, \citenamefont {Stobbe}, \citenamefont
  {Sotomayor-Torres},\ and\ \citenamefont {García}}]{Florez2022}%
  \BibitemOpen
  \bibfield  {author} {\bibinfo {author} {\bibfnamefont {O.}~\bibnamefont
  {Florez}}, \bibinfo {author} {\bibfnamefont {G.}~\bibnamefont {Arregui}},
  \bibinfo {author} {\bibfnamefont {M.}~\bibnamefont {Albrechtsen}}, \bibinfo
  {author} {\bibfnamefont {R.~C.}\ \bibnamefont {Ng}}, \bibinfo {author}
  {\bibfnamefont {J.}~\bibnamefont {Gomis-Bresco}}, \bibinfo {author}
  {\bibfnamefont {S.}~\bibnamefont {Stobbe}}, \bibinfo {author} {\bibfnamefont
  {C.~M.}\ \bibnamefont {Sotomayor-Torres}}, \ and\ \bibinfo {author}
  {\bibfnamefont {P.~D.}\ \bibnamefont {García}},\ }\href {\doibase Accepted}
  {\bibfield  {journal} {\bibinfo  {journal} {Nature Nanotechnology}\ }
  (\bibinfo {year} {2022}),\ Accepted},\ \Eprint
  {http://arxiv.org/abs/2202.02166} {arXiv:2202.02166} \BibitemShut {NoStop}%
\bibitem [{\citenamefont {Fuhrmann}\ \emph {et~al.}(2011)\citenamefont
  {Fuhrmann}, \citenamefont {Thon}, \citenamefont {Kim}, \citenamefont
  {Bouwmeester}, \citenamefont {Petroff}, \citenamefont {Wixforth},\ and\
  \citenamefont {Krenner}}]{Fuhrmann2011}%
  \BibitemOpen
  \bibfield  {author} {\bibinfo {author} {\bibfnamefont {D.~A.}\ \bibnamefont
  {Fuhrmann}}, \bibinfo {author} {\bibfnamefont {S.~M.}\ \bibnamefont {Thon}},
  \bibinfo {author} {\bibfnamefont {H.}~\bibnamefont {Kim}}, \bibinfo {author}
  {\bibfnamefont {D.}~\bibnamefont {Bouwmeester}}, \bibinfo {author}
  {\bibfnamefont {P.~M.}\ \bibnamefont {Petroff}}, \bibinfo {author}
  {\bibfnamefont {A.}~\bibnamefont {Wixforth}}, \ and\ \bibinfo {author}
  {\bibfnamefont {H.~J.}\ \bibnamefont {Krenner}},\ }\href {\doibase
  10.1038/nphoton.2011.208} {\bibfield  {journal} {\bibinfo  {journal} {Nature
  Photonics}\ }\textbf {\bibinfo {volume} {5}},\ \bibinfo {pages} {605}
  (\bibinfo {year} {2011})}\BibitemShut {NoStop}%
\bibitem [{\citenamefont {Mahboob}\ \emph {et~al.}(2013)\citenamefont
  {Mahboob}, \citenamefont {Nishiguchi}, \citenamefont {Fujiwara},\ and\
  \citenamefont {Yamaguchi}}]{Mahboob2013}%
  \BibitemOpen
  \bibfield  {author} {\bibinfo {author} {\bibfnamefont {I.}~\bibnamefont
  {Mahboob}}, \bibinfo {author} {\bibfnamefont {K.}~\bibnamefont {Nishiguchi}},
  \bibinfo {author} {\bibfnamefont {A.}~\bibnamefont {Fujiwara}}, \ and\
  \bibinfo {author} {\bibfnamefont {H.}~\bibnamefont {Yamaguchi}},\ }\href
  {\doibase 10.1103/PhysRevLett.110.127202} {\bibfield  {journal} {\bibinfo
  {journal} {Phys. Rev. Lett.}\ }\textbf {\bibinfo {volume} {110}},\ \bibinfo
  {pages} {127202} (\bibinfo {year} {2013})}\BibitemShut {NoStop}%
\bibitem [{\citenamefont {Balram}\ \emph {et~al.}(2016)\citenamefont {Balram},
  \citenamefont {Davan{\c{c}}o}, \citenamefont {Song},\ and\ \citenamefont
  {Srinivasan}}]{Balram2016}%
  \BibitemOpen
  \bibfield  {author} {\bibinfo {author} {\bibfnamefont {K.~C.}\ \bibnamefont
  {Balram}}, \bibinfo {author} {\bibfnamefont {M.~I.}\ \bibnamefont
  {Davan{\c{c}}o}}, \bibinfo {author} {\bibfnamefont {J.~D.}\ \bibnamefont
  {Song}}, \ and\ \bibinfo {author} {\bibfnamefont {K.}~\bibnamefont
  {Srinivasan}},\ }\href {\doibase 10.1038/nphoton.2016.46} {\bibfield
  {journal} {\bibinfo  {journal} {Nature Photonics}\ }\textbf {\bibinfo
  {volume} {10}},\ \bibinfo {pages} {346} (\bibinfo {year} {2016})}\BibitemShut
  {NoStop}%
\bibitem [{\citenamefont {Korovin}\ \emph {et~al.}(2019)\citenamefont
  {Korovin}, \citenamefont {Pennec}, \citenamefont {Stocchi}, \citenamefont
  {Mencarelli}, \citenamefont {Pierantoni}, \citenamefont {Makkonen},
  \citenamefont {Ahopelto},\ and\ \citenamefont
  {Djafari-Rouhani}}]{Korovin2019}%
  \BibitemOpen
  \bibfield  {author} {\bibinfo {author} {\bibfnamefont {A.}~\bibnamefont
  {Korovin}}, \bibinfo {author} {\bibfnamefont {Y.}~\bibnamefont {Pennec}},
  \bibinfo {author} {\bibfnamefont {M.}~\bibnamefont {Stocchi}}, \bibinfo
  {author} {\bibfnamefont {D.}~\bibnamefont {Mencarelli}}, \bibinfo {author}
  {\bibfnamefont {L.}~\bibnamefont {Pierantoni}}, \bibinfo {author}
  {\bibfnamefont {T.}~\bibnamefont {Makkonen}}, \bibinfo {author}
  {\bibfnamefont {J.}~\bibnamefont {Ahopelto}}, \ and\ \bibinfo {author}
  {\bibfnamefont {B.}~\bibnamefont {Djafari-Rouhani}},\ }\href {\doibase
  10.1088/1361-6463/ab2297} {\bibfield  {journal} {\bibinfo  {journal} {Journal
  of Physics D: Applied Physics}\ }\textbf {\bibinfo {volume} {52}} (\bibinfo
  {year} {2019}),\ 10.1088/1361-6463/ab2297}\BibitemShut {NoStop}%
\bibitem [{\citenamefont {Kuznetsov}\ \emph {et~al.}(2021)\citenamefont
  {Kuznetsov}, \citenamefont {Machado}, \citenamefont {Biermann},\ and\
  \citenamefont {Santos}}]{Kuznetsov2021}%
  \BibitemOpen
  \bibfield  {author} {\bibinfo {author} {\bibfnamefont {A.~S.}\ \bibnamefont
  {Kuznetsov}}, \bibinfo {author} {\bibfnamefont {D.~H.~O.}\ \bibnamefont
  {Machado}}, \bibinfo {author} {\bibfnamefont {K.}~\bibnamefont {Biermann}}, \
  and\ \bibinfo {author} {\bibfnamefont {P.~V.}\ \bibnamefont {Santos}},\
  }\href {\doibase 10.1103/PhysRevX.11.021020} {\bibfield  {journal} {\bibinfo
  {journal} {Phys. Rev. X}\ }\textbf {\bibinfo {volume} {11}},\ \bibinfo
  {pages} {021020} (\bibinfo {year} {2021})}\BibitemShut {NoStop}%
\bibitem [{\citenamefont {Munk}\ \emph {et~al.}(2019)\citenamefont {Munk},
  \citenamefont {Katzman}, \citenamefont {Hen}, \citenamefont {Priel},
  \citenamefont {Feldberg}, \citenamefont {Sharabani}, \citenamefont {Levy},
  \citenamefont {Bergman},\ and\ \citenamefont {Zadok}}]{munk2019surface}%
  \BibitemOpen
  \bibfield  {author} {\bibinfo {author} {\bibfnamefont {D.}~\bibnamefont
  {Munk}}, \bibinfo {author} {\bibfnamefont {M.}~\bibnamefont {Katzman}},
  \bibinfo {author} {\bibfnamefont {M.}~\bibnamefont {Hen}}, \bibinfo {author}
  {\bibfnamefont {M.}~\bibnamefont {Priel}}, \bibinfo {author} {\bibfnamefont
  {M.}~\bibnamefont {Feldberg}}, \bibinfo {author} {\bibfnamefont
  {T.}~\bibnamefont {Sharabani}}, \bibinfo {author} {\bibfnamefont
  {S.}~\bibnamefont {Levy}}, \bibinfo {author} {\bibfnamefont {A.}~\bibnamefont
  {Bergman}}, \ and\ \bibinfo {author} {\bibfnamefont {A.}~\bibnamefont
  {Zadok}},\ }\href@noop {} {\bibfield  {journal} {\bibinfo  {journal} {Nature
  communications}\ }\textbf {\bibinfo {volume} {10}},\ \bibinfo {pages} {1}
  (\bibinfo {year} {2019})}\BibitemShut {NoStop}%
\bibitem [{\citenamefont {Grudinin}\ \emph {et~al.}(2010)\citenamefont
  {Grudinin}, \citenamefont {Lee}, \citenamefont {Painter},\ and\ \citenamefont
  {Vahala}}]{Grudinin2010}%
  \BibitemOpen
  \bibfield  {author} {\bibinfo {author} {\bibfnamefont {I.~S.}\ \bibnamefont
  {Grudinin}}, \bibinfo {author} {\bibfnamefont {H.}~\bibnamefont {Lee}},
  \bibinfo {author} {\bibfnamefont {O.}~\bibnamefont {Painter}}, \ and\
  \bibinfo {author} {\bibfnamefont {K.~J.}\ \bibnamefont {Vahala}},\ }\href
  {\doibase 10.1103/PhysRevLett.104.083901} {\bibfield  {journal} {\bibinfo
  {journal} {Phys. Rev. Lett.}\ }\textbf {\bibinfo {volume} {104}},\ \bibinfo
  {pages} {83901} (\bibinfo {year} {2010})}\BibitemShut {NoStop}%
\bibitem [{\citenamefont {Burek}\ \emph {et~al.}(2016)\citenamefont {Burek},
  \citenamefont {Cohen}, \citenamefont {Meenehan}, \citenamefont {El-Sawah},
  \citenamefont {Chia}, \citenamefont {Ruelle}, \citenamefont {Meesala},
  \citenamefont {Rochman}, \citenamefont {Atikian}, \citenamefont {Markham},
  \citenamefont {Twitchen}, \citenamefont {Lukin}, \citenamefont {Painter},\
  and\ \citenamefont {Lon\v{c}ar}}]{Burek:16}%
  \BibitemOpen
  \bibfield  {author} {\bibinfo {author} {\bibfnamefont {M.~J.}\ \bibnamefont
  {Burek}}, \bibinfo {author} {\bibfnamefont {J.~D.}\ \bibnamefont {Cohen}},
  \bibinfo {author} {\bibfnamefont {S.~M.}\ \bibnamefont {Meenehan}}, \bibinfo
  {author} {\bibfnamefont {N.}~\bibnamefont {El-Sawah}}, \bibinfo {author}
  {\bibfnamefont {C.}~\bibnamefont {Chia}}, \bibinfo {author} {\bibfnamefont
  {T.}~\bibnamefont {Ruelle}}, \bibinfo {author} {\bibfnamefont
  {S.}~\bibnamefont {Meesala}}, \bibinfo {author} {\bibfnamefont
  {J.}~\bibnamefont {Rochman}}, \bibinfo {author} {\bibfnamefont {H.~A.}\
  \bibnamefont {Atikian}}, \bibinfo {author} {\bibfnamefont {M.}~\bibnamefont
  {Markham}}, \bibinfo {author} {\bibfnamefont {D.~J.}\ \bibnamefont
  {Twitchen}}, \bibinfo {author} {\bibfnamefont {M.~D.}\ \bibnamefont {Lukin}},
  \bibinfo {author} {\bibfnamefont {O.}~\bibnamefont {Painter}}, \ and\
  \bibinfo {author} {\bibfnamefont {M.}~\bibnamefont {Lon\v{c}ar}},\ }\href
  {\doibase 10.1364/OPTICA.3.001404} {\bibfield  {journal} {\bibinfo  {journal}
  {Optica}\ }\textbf {\bibinfo {volume} {3}},\ \bibinfo {pages} {1404}
  (\bibinfo {year} {2016})}\BibitemShut {NoStop}%
\bibitem [{\citenamefont {Navarro-Urrios}\ \emph {et~al.}(2015)\citenamefont
  {Navarro-Urrios}, \citenamefont {Capuj}, \citenamefont {Gomis-Bresco},
  \citenamefont {Alzina}, \citenamefont {Pitanti}, \citenamefont {Griol},
  \citenamefont {Mart{\'{i}}nez},\ and\ \citenamefont
  {{Sotomayor-Torres}}}]{NavarroUrrios2015}%
  \BibitemOpen
  \bibfield  {author} {\bibinfo {author} {\bibfnamefont {D.}~\bibnamefont
  {Navarro-Urrios}}, \bibinfo {author} {\bibfnamefont {N.~E.}\ \bibnamefont
  {Capuj}}, \bibinfo {author} {\bibfnamefont {J.}~\bibnamefont {Gomis-Bresco}},
  \bibinfo {author} {\bibfnamefont {F.}~\bibnamefont {Alzina}}, \bibinfo
  {author} {\bibfnamefont {A.}~\bibnamefont {Pitanti}}, \bibinfo {author}
  {\bibfnamefont {A.}~\bibnamefont {Griol}}, \bibinfo {author} {\bibfnamefont
  {A.}~\bibnamefont {Mart{\'{i}}nez}}, \ and\ \bibinfo {author} {\bibfnamefont
  {C.~M.}\ \bibnamefont {{Sotomayor-Torres}}},\ }\href {\doibase
  10.1038/srep15733} {\bibfield  {journal} {\bibinfo  {journal} {Scientific
  Reports}\ }\textbf {\bibinfo {volume} {5}},\ \bibinfo {pages} {15733}
  (\bibinfo {year} {2015})}\BibitemShut {NoStop}%
\bibitem [{\citenamefont {Ghorbel}\ \emph
  {et~al.}(2019{\natexlab{a}})\citenamefont {Ghorbel}, \citenamefont {Swiadek},
  \citenamefont {Zhu}, \citenamefont {Dolfi}, \citenamefont {Lehoucq},
  \citenamefont {Martin}, \citenamefont {Moille}, \citenamefont {Morvan},
  \citenamefont {Braive}, \citenamefont {Combri{\'e}} \emph
  {et~al.}}]{ghorbel2019optomechanical}%
  \BibitemOpen
  \bibfield  {author} {\bibinfo {author} {\bibfnamefont {I.}~\bibnamefont
  {Ghorbel}}, \bibinfo {author} {\bibfnamefont {F.}~\bibnamefont {Swiadek}},
  \bibinfo {author} {\bibfnamefont {R.}~\bibnamefont {Zhu}}, \bibinfo {author}
  {\bibfnamefont {D.}~\bibnamefont {Dolfi}}, \bibinfo {author} {\bibfnamefont
  {G.}~\bibnamefont {Lehoucq}}, \bibinfo {author} {\bibfnamefont
  {A.}~\bibnamefont {Martin}}, \bibinfo {author} {\bibfnamefont
  {G.}~\bibnamefont {Moille}}, \bibinfo {author} {\bibfnamefont
  {L.}~\bibnamefont {Morvan}}, \bibinfo {author} {\bibfnamefont
  {R.}~\bibnamefont {Braive}}, \bibinfo {author} {\bibfnamefont
  {S.}~\bibnamefont {Combri{\'e}}},  \emph {et~al.},\ }\href@noop {} {\bibfield
   {journal} {\bibinfo  {journal} {APL Photonics}\ }\textbf {\bibinfo {volume}
  {4}},\ \bibinfo {pages} {116103} (\bibinfo {year}
  {2019}{\natexlab{a}})}\BibitemShut {NoStop}%
\bibitem [{\citenamefont {Mercad{\'e}}\ \emph {et~al.}(2021)\citenamefont
  {Mercad{\'e}}, \citenamefont {Pelka}, \citenamefont {Burgwal}, \citenamefont
  {Xuereb}, \citenamefont {Mart{\'\i}nez},\ and\ \citenamefont
  {Verhagen}}]{mercade2021floquet}%
  \BibitemOpen
  \bibfield  {author} {\bibinfo {author} {\bibfnamefont {L.}~\bibnamefont
  {Mercad{\'e}}}, \bibinfo {author} {\bibfnamefont {K.}~\bibnamefont {Pelka}},
  \bibinfo {author} {\bibfnamefont {R.}~\bibnamefont {Burgwal}}, \bibinfo
  {author} {\bibfnamefont {A.}~\bibnamefont {Xuereb}}, \bibinfo {author}
  {\bibfnamefont {A.}~\bibnamefont {Mart{\'\i}nez}}, \ and\ \bibinfo {author}
  {\bibfnamefont {E.}~\bibnamefont {Verhagen}},\ }\href@noop {} {\bibfield
  {journal} {\bibinfo  {journal} {Physical Review Letters}\ }\textbf {\bibinfo
  {volume} {127}},\ \bibinfo {pages} {073601} (\bibinfo {year}
  {2021})}\BibitemShut {NoStop}%
\bibitem [{\citenamefont {Fang}\ \emph {et~al.}(2016)\citenamefont {Fang},
  \citenamefont {Matheny}, \citenamefont {Luan},\ and\ \citenamefont
  {Painter}}]{Fang2016}%
  \BibitemOpen
  \bibfield  {author} {\bibinfo {author} {\bibfnamefont {K.}~\bibnamefont
  {Fang}}, \bibinfo {author} {\bibfnamefont {M.~H.}\ \bibnamefont {Matheny}},
  \bibinfo {author} {\bibfnamefont {X.}~\bibnamefont {Luan}}, \ and\ \bibinfo
  {author} {\bibfnamefont {O.}~\bibnamefont {Painter}},\ }\href {\doibase
  10.1038/nphoton.2016.107} {\bibfield  {journal} {\bibinfo  {journal} {Nature
  Photonics}\ }\textbf {\bibinfo {volume} {10}},\ \bibinfo {pages} {489}
  (\bibinfo {year} {2016})}\BibitemShut {NoStop}%
\bibitem [{\citenamefont {Guo}\ \emph {et~al.}(2019)\citenamefont {Guo},
  \citenamefont {Norte},\ and\ \citenamefont {Gr\"oblacher}}]{Guo2019}%
  \BibitemOpen
  \bibfield  {author} {\bibinfo {author} {\bibfnamefont {J.}~\bibnamefont
  {Guo}}, \bibinfo {author} {\bibfnamefont {R.}~\bibnamefont {Norte}}, \ and\
  \bibinfo {author} {\bibfnamefont {S.}~\bibnamefont {Gr\"oblacher}},\ }\href
  {\doibase 10.1103/PhysRevLett.123.223602} {\bibfield  {journal} {\bibinfo
  {journal} {Phys. Rev. Lett.}\ }\textbf {\bibinfo {volume} {123}},\ \bibinfo
  {pages} {223602} (\bibinfo {year} {2019})}\BibitemShut {NoStop}%
\bibitem [{\citenamefont {Mayor}\ \emph {et~al.}(2021)\citenamefont {Mayor},
  \citenamefont {Jiang}, \citenamefont {Sarabalis}, \citenamefont {McKenna},
  \citenamefont {Witmer},\ and\ \citenamefont {Safavi-Naeini}}]{Mayor2021}%
  \BibitemOpen
  \bibfield  {author} {\bibinfo {author} {\bibfnamefont {F.~M.}\ \bibnamefont
  {Mayor}}, \bibinfo {author} {\bibfnamefont {W.}~\bibnamefont {Jiang}},
  \bibinfo {author} {\bibfnamefont {C.~J.}\ \bibnamefont {Sarabalis}}, \bibinfo
  {author} {\bibfnamefont {T.~P.}\ \bibnamefont {McKenna}}, \bibinfo {author}
  {\bibfnamefont {J.~D.}\ \bibnamefont {Witmer}}, \ and\ \bibinfo {author}
  {\bibfnamefont {A.~H.}\ \bibnamefont {Safavi-Naeini}},\ }\href {\doibase
  10.1103/PhysRevApplied.15.014039} {\bibfield  {journal} {\bibinfo  {journal}
  {Phys. Rev. Applied}\ }\textbf {\bibinfo {volume} {15}},\ \bibinfo {pages}
  {014039} (\bibinfo {year} {2021})}\BibitemShut {NoStop}%
\bibitem [{\citenamefont {Patel}\ \emph {et~al.}(2018)\citenamefont {Patel},
  \citenamefont {Wang}, \citenamefont {Jiang}, \citenamefont {Sarabalis},
  \citenamefont {Hill},\ and\ \citenamefont {Safavi-Naeini}}]{Patel2018}%
  \BibitemOpen
  \bibfield  {author} {\bibinfo {author} {\bibfnamefont {R.~N.}\ \bibnamefont
  {Patel}}, \bibinfo {author} {\bibfnamefont {Z.}~\bibnamefont {Wang}},
  \bibinfo {author} {\bibfnamefont {W.}~\bibnamefont {Jiang}}, \bibinfo
  {author} {\bibfnamefont {C.~J.}\ \bibnamefont {Sarabalis}}, \bibinfo {author}
  {\bibfnamefont {J.~T.}\ \bibnamefont {Hill}}, \ and\ \bibinfo {author}
  {\bibfnamefont {A.~H.}\ \bibnamefont {Safavi-Naeini}},\ }\href {\doibase
  10.1103/PhysRevLett.121.040501} {\bibfield  {journal} {\bibinfo  {journal}
  {Phys. Rev. Lett.}\ }\textbf {\bibinfo {volume} {121}},\ \bibinfo {pages}
  {40501} (\bibinfo {year} {2018})}\BibitemShut {NoStop}%
\bibitem [{\citenamefont {Zivari}\ \emph
  {et~al.}(2022{\natexlab{b}})\citenamefont {Zivari}, \citenamefont {Stockill},
  \citenamefont {Fiaschi},\ and\ \citenamefont
  {Gr{\"o}blacher}}]{zivari2022non}%
  \BibitemOpen
  \bibfield  {author} {\bibinfo {author} {\bibfnamefont {A.}~\bibnamefont
  {Zivari}}, \bibinfo {author} {\bibfnamefont {R.}~\bibnamefont {Stockill}},
  \bibinfo {author} {\bibfnamefont {N.}~\bibnamefont {Fiaschi}}, \ and\
  \bibinfo {author} {\bibfnamefont {S.}~\bibnamefont {Gr{\"o}blacher}},\
  }\href@noop {} {\bibfield  {journal} {\bibinfo  {journal} {Nature Physics}\
  ,\ \bibinfo {pages} {1}} (\bibinfo {year} {2022}{\natexlab{b}})}\BibitemShut
  {NoStop}%
\bibitem [{\citenamefont {Wen}\ \emph {et~al.}(2008)\citenamefont {Wen},
  \citenamefont {David}, \citenamefont {Checoury}, \citenamefont {Kurdi},\ and\
  \citenamefont {Boucaud}}]{Wen:08}%
  \BibitemOpen
  \bibfield  {author} {\bibinfo {author} {\bibfnamefont {F.}~\bibnamefont
  {Wen}}, \bibinfo {author} {\bibfnamefont {S.}~\bibnamefont {David}}, \bibinfo
  {author} {\bibfnamefont {X.}~\bibnamefont {Checoury}}, \bibinfo {author}
  {\bibfnamefont {M.~E.}\ \bibnamefont {Kurdi}}, \ and\ \bibinfo {author}
  {\bibfnamefont {P.}~\bibnamefont {Boucaud}},\ }\href {\doibase
  10.1364/OE.16.012278} {\bibfield  {journal} {\bibinfo  {journal} {Opt.
  Express}\ }\textbf {\bibinfo {volume} {16}},\ \bibinfo {pages} {12278}
  (\bibinfo {year} {2008})}\BibitemShut {NoStop}%
\bibitem [{\citenamefont {Gomis-Bresco}\ \emph {et~al.}(2014)\citenamefont
  {Gomis-Bresco}, \citenamefont {Navarro-Urrios}, \citenamefont {Oudich},
  \citenamefont {El-Jallal}, \citenamefont {Griol}, \citenamefont {Puerto},
  \citenamefont {Chavez}, \citenamefont {Pennec}, \citenamefont
  {Djafari-Rouhani}, \citenamefont {Alzina}, \citenamefont {Mart{\'{i}}nez},\
  and\ \citenamefont {Sotomayor-Torres}}]{GomisBresco2014}%
  \BibitemOpen
  \bibfield  {author} {\bibinfo {author} {\bibfnamefont {J.}~\bibnamefont
  {Gomis-Bresco}}, \bibinfo {author} {\bibfnamefont {D.}~\bibnamefont
  {Navarro-Urrios}}, \bibinfo {author} {\bibfnamefont {M.}~\bibnamefont
  {Oudich}}, \bibinfo {author} {\bibfnamefont {S.}~\bibnamefont {El-Jallal}},
  \bibinfo {author} {\bibfnamefont {A.}~\bibnamefont {Griol}}, \bibinfo
  {author} {\bibfnamefont {D.}~\bibnamefont {Puerto}}, \bibinfo {author}
  {\bibfnamefont {E.}~\bibnamefont {Chavez}}, \bibinfo {author} {\bibfnamefont
  {Y.}~\bibnamefont {Pennec}}, \bibinfo {author} {\bibfnamefont
  {B.}~\bibnamefont {Djafari-Rouhani}}, \bibinfo {author} {\bibfnamefont
  {F.}~\bibnamefont {Alzina}}, \bibinfo {author} {\bibfnamefont
  {A.}~\bibnamefont {Mart{\'{i}}nez}}, \ and\ \bibinfo {author} {\bibfnamefont
  {C.~M.}\ \bibnamefont {Sotomayor-Torres}},\ }\href {\doibase
  10.1038/ncomms5452} {\bibfield  {journal} {\bibinfo  {journal} {Nature
  Communications}\ }\textbf {\bibinfo {volume} {5}},\ \bibinfo {pages} {4452}
  (\bibinfo {year} {2014})}\BibitemShut {NoStop}%
\bibitem [{\citenamefont {Garc\'{\i}a}\ \emph {et~al.}(2017)\citenamefont
  {Garc\'{\i}a}, \citenamefont {Bericat-Vadell}, \citenamefont {Arregui},
  \citenamefont {Navarro-Urrios}, \citenamefont {Colombano}, \citenamefont
  {Alzina},\ and\ \citenamefont {Sotomayor-Torres}}]{PhysRevB.95.115129}%
  \BibitemOpen
  \bibfield  {author} {\bibinfo {author} {\bibfnamefont {P.~D.}\ \bibnamefont
  {Garc\'{\i}a}}, \bibinfo {author} {\bibfnamefont {R.}~\bibnamefont
  {Bericat-Vadell}}, \bibinfo {author} {\bibfnamefont {G.}~\bibnamefont
  {Arregui}}, \bibinfo {author} {\bibfnamefont {D.}~\bibnamefont
  {Navarro-Urrios}}, \bibinfo {author} {\bibfnamefont {M.}~\bibnamefont
  {Colombano}}, \bibinfo {author} {\bibfnamefont {F.}~\bibnamefont {Alzina}}, \
  and\ \bibinfo {author} {\bibfnamefont {C.~M.}\ \bibnamefont
  {Sotomayor-Torres}},\ }\href {\doibase 10.1103/PhysRevB.95.115129} {\bibfield
   {journal} {\bibinfo  {journal} {Phys. Rev. B}\ }\textbf {\bibinfo {volume}
  {95}},\ \bibinfo {pages} {115129} (\bibinfo {year} {2017})}\BibitemShut
  {NoStop}%
\bibitem [{\citenamefont {Arregui}\ \emph {et~al.}(2018)\citenamefont
  {Arregui}, \citenamefont {Navarro-Urrios}, \citenamefont {Kehagias},
  \citenamefont {Torres},\ and\ \citenamefont {Garc\'{\i}a}}]{ArreguiPRB2018}%
  \BibitemOpen
  \bibfield  {author} {\bibinfo {author} {\bibfnamefont {G.}~\bibnamefont
  {Arregui}}, \bibinfo {author} {\bibfnamefont {D.}~\bibnamefont
  {Navarro-Urrios}}, \bibinfo {author} {\bibfnamefont {N.}~\bibnamefont
  {Kehagias}}, \bibinfo {author} {\bibfnamefont {C.~M.~S.}\ \bibnamefont
  {Torres}}, \ and\ \bibinfo {author} {\bibfnamefont {P.~D.}\ \bibnamefont
  {Garc\'{\i}a}},\ }\href {\doibase 10.1103/PhysRevB.98.180202} {\bibfield
  {journal} {\bibinfo  {journal} {Phys. Rev. B}\ }\textbf {\bibinfo {volume}
  {98}},\ \bibinfo {pages} {180202} (\bibinfo {year} {2018})}\BibitemShut
  {NoStop}%
\bibitem [{\citenamefont {Arregui}\ \emph {et~al.}(2021)\citenamefont
  {Arregui}, \citenamefont {Ng}, \citenamefont {Albrechtsen}, \citenamefont
  {Stobbe}, \citenamefont {Sotomayor-Torres},\ and\ \citenamefont
  {Garc{\'{i}}a}}]{Arregui2021arXiv}%
  \BibitemOpen
  \bibfield  {author} {\bibinfo {author} {\bibfnamefont {G.}~\bibnamefont
  {Arregui}}, \bibinfo {author} {\bibfnamefont {R.~C.}\ \bibnamefont {Ng}},
  \bibinfo {author} {\bibfnamefont {M.}~\bibnamefont {Albrechtsen}}, \bibinfo
  {author} {\bibfnamefont {S.}~\bibnamefont {Stobbe}}, \bibinfo {author}
  {\bibfnamefont {C.~M.}\ \bibnamefont {Sotomayor-Torres}}, \ and\ \bibinfo
  {author} {\bibfnamefont {P.~D.}\ \bibnamefont {Garc{\'{i}}a}},\ }\href@noop
  {} {\enquote {\bibinfo {title} {{Cavity optomechanics with Anderson-localized
  optical modes}},}\ } (\bibinfo {year} {2021}),\ \Eprint
  {http://arxiv.org/abs/2110.11005} {arXiv:2110.11005 [physics.optics]}
  \BibitemShut {NoStop}%
\bibitem [{\citenamefont {Anderson}(1958)}]{PhysRev.109.1492}%
  \BibitemOpen
  \bibfield  {author} {\bibinfo {author} {\bibfnamefont {P.~W.}\ \bibnamefont
  {Anderson}},\ }\href {\doibase 10.1103/PhysRev.109.1492} {\bibfield
  {journal} {\bibinfo  {journal} {Phys. Rev.}\ }\textbf {\bibinfo {volume}
  {109}},\ \bibinfo {pages} {1492} (\bibinfo {year} {1958})}\BibitemShut
  {NoStop}%
\bibitem [{\citenamefont {{Le Thomas}}\ \emph {et~al.}(2009)\citenamefont {{Le
  Thomas}}, \citenamefont {Zhang}, \citenamefont {J{\'{a}}gersk{\'{a}}},
  \citenamefont {Zabelin}, \citenamefont {Houdr{\'{e}}}, \citenamefont
  {Sagnes},\ and\ \citenamefont {Talneau}}]{LeThomas2009}%
  \BibitemOpen
  \bibfield  {author} {\bibinfo {author} {\bibfnamefont {N.}~\bibnamefont {{Le
  Thomas}}}, \bibinfo {author} {\bibfnamefont {H.}~\bibnamefont {Zhang}},
  \bibinfo {author} {\bibfnamefont {J.}~\bibnamefont {J{\'{a}}gersk{\'{a}}}},
  \bibinfo {author} {\bibfnamefont {V.}~\bibnamefont {Zabelin}}, \bibinfo
  {author} {\bibfnamefont {R.}~\bibnamefont {Houdr{\'{e}}}}, \bibinfo {author}
  {\bibfnamefont {I.}~\bibnamefont {Sagnes}}, \ and\ \bibinfo {author}
  {\bibfnamefont {A.}~\bibnamefont {Talneau}},\ }\href {\doibase
  10.1103/PhysRevB.80.125332} {\bibfield  {journal} {\bibinfo  {journal} {Phys.
  Rev. B}\ }\textbf {\bibinfo {volume} {80}},\ \bibinfo {pages} {125332}
  (\bibinfo {year} {2009})}\BibitemShut {NoStop}%
\bibitem [{\citenamefont {Sapienza}\ \emph {et~al.}(2010)\citenamefont
  {Sapienza}, \citenamefont {Thyrrestrup}, \citenamefont {Stobbe},
  \citenamefont {Garcia}, \citenamefont {Smolka},\ and\ \citenamefont
  {Lodahl}}]{Sapienza2010}%
  \BibitemOpen
  \bibfield  {author} {\bibinfo {author} {\bibfnamefont {L.}~\bibnamefont
  {Sapienza}}, \bibinfo {author} {\bibfnamefont {H.}~\bibnamefont
  {Thyrrestrup}}, \bibinfo {author} {\bibfnamefont {S.}~\bibnamefont {Stobbe}},
  \bibinfo {author} {\bibfnamefont {P.~D.}\ \bibnamefont {Garcia}}, \bibinfo
  {author} {\bibfnamefont {S.}~\bibnamefont {Smolka}}, \ and\ \bibinfo {author}
  {\bibfnamefont {P.}~\bibnamefont {Lodahl}},\ }in\ \href {\doibase
  10.1364/QELS.2010.QPDA7} {\emph {\bibinfo {booktitle} {Conference on Lasers
  and Electro-Optics 2010}}}\ (\bibinfo  {publisher} {Optical Society of
  America},\ \bibinfo {year} {2010})\ p.\ \bibinfo {pages} {QPDA7}\BibitemShut
  {NoStop}%
\bibitem [{\citenamefont {Eichenfield}\ \emph {et~al.}(2009)\citenamefont
  {Eichenfield}, \citenamefont {Chan}, \citenamefont {Camacho}, \citenamefont
  {Vahala},\ and\ \citenamefont {Painter}}]{Eichenfield2009}%
  \BibitemOpen
  \bibfield  {author} {\bibinfo {author} {\bibfnamefont {M.}~\bibnamefont
  {Eichenfield}}, \bibinfo {author} {\bibfnamefont {J.}~\bibnamefont {Chan}},
  \bibinfo {author} {\bibfnamefont {R.~M.}\ \bibnamefont {Camacho}}, \bibinfo
  {author} {\bibfnamefont {K.~J.}\ \bibnamefont {Vahala}}, \ and\ \bibinfo
  {author} {\bibfnamefont {O.}~\bibnamefont {Painter}},\ }\href {\doibase
  10.1038/nature08524} {\bibfield  {journal} {\bibinfo  {journal} {Nature}\
  }\textbf {\bibinfo {volume} {462}},\ \bibinfo {pages} {78} (\bibinfo {year}
  {2009})}\BibitemShut {NoStop}%
\bibitem [{\citenamefont {Mitchell}\ \emph {et~al.}(2014)\citenamefont
  {Mitchell}, \citenamefont {Hryciw},\ and\ \citenamefont
  {Barclay}}]{Mitchell2014}%
  \BibitemOpen
  \bibfield  {author} {\bibinfo {author} {\bibfnamefont {M.}~\bibnamefont
  {Mitchell}}, \bibinfo {author} {\bibfnamefont {A.~C.}\ \bibnamefont
  {Hryciw}}, \ and\ \bibinfo {author} {\bibfnamefont {P.~E.}\ \bibnamefont
  {Barclay}},\ }\href {\doibase 10.1063/1.4870999} {\bibfield  {journal}
  {\bibinfo  {journal} {Applied Physics Letters}\ }\textbf {\bibinfo {volume}
  {104}},\ \bibinfo {pages} {141104} (\bibinfo {year} {2014})},\ \Eprint
  {http://arxiv.org/abs/https://doi.org/10.1063/1.4870999}
  {https://doi.org/10.1063/1.4870999} \BibitemShut {NoStop}%
\bibitem [{\citenamefont {Ghorbel}\ \emph
  {et~al.}(2019{\natexlab{b}})\citenamefont {Ghorbel}, \citenamefont {Swiadek},
  \citenamefont {Zhu}, \citenamefont {Dolfi}, \citenamefont {Lehoucq},
  \citenamefont {Martin}, \citenamefont {Moille}, \citenamefont {Morvan},
  \citenamefont {Braive}, \citenamefont {Combrié},\ and\ \citenamefont
  {De~Rossi}}]{Ghorbel2019}%
  \BibitemOpen
  \bibfield  {author} {\bibinfo {author} {\bibfnamefont {I.}~\bibnamefont
  {Ghorbel}}, \bibinfo {author} {\bibfnamefont {F.}~\bibnamefont {Swiadek}},
  \bibinfo {author} {\bibfnamefont {R.}~\bibnamefont {Zhu}}, \bibinfo {author}
  {\bibfnamefont {D.}~\bibnamefont {Dolfi}}, \bibinfo {author} {\bibfnamefont
  {G.}~\bibnamefont {Lehoucq}}, \bibinfo {author} {\bibfnamefont
  {A.}~\bibnamefont {Martin}}, \bibinfo {author} {\bibfnamefont
  {G.}~\bibnamefont {Moille}}, \bibinfo {author} {\bibfnamefont
  {L.}~\bibnamefont {Morvan}}, \bibinfo {author} {\bibfnamefont
  {R.}~\bibnamefont {Braive}}, \bibinfo {author} {\bibfnamefont
  {S.}~\bibnamefont {Combrié}}, \ and\ \bibinfo {author} {\bibfnamefont
  {A.}~\bibnamefont {De~Rossi}},\ }\href {\doibase 10.1063/1.5121774}
  {\bibfield  {journal} {\bibinfo  {journal} {APL Photonics}\ }\textbf
  {\bibinfo {volume} {4}},\ \bibinfo {pages} {116103} (\bibinfo {year}
  {2019}{\natexlab{b}})},\ \Eprint
  {http://arxiv.org/abs/https://doi.org/10.1063/1.5121774}
  {https://doi.org/10.1063/1.5121774} \BibitemShut {NoStop}%
\bibitem [{\citenamefont {Topolancik}\ \emph {et~al.}(2007)\citenamefont
  {Topolancik}, \citenamefont {Vollmer},\ and\ \citenamefont
  {Ilic}}]{Topolancik2007}%
  \BibitemOpen
  \bibfield  {author} {\bibinfo {author} {\bibfnamefont {J.}~\bibnamefont
  {Topolancik}}, \bibinfo {author} {\bibfnamefont {F.}~\bibnamefont {Vollmer}},
  \ and\ \bibinfo {author} {\bibfnamefont {B.}~\bibnamefont {Ilic}},\ }\href
  {\doibase 10.1063/1.2809614} {\bibfield  {journal} {\bibinfo  {journal}
  {Applied Physics Letters}\ }\textbf {\bibinfo {volume} {91}},\ \bibinfo
  {pages} {201102} (\bibinfo {year} {2007})},\ \Eprint
  {http://arxiv.org/abs/https://doi.org/10.1063/1.2809614}
  {https://doi.org/10.1063/1.2809614} \BibitemShut {NoStop}%
\bibitem [{\citenamefont {Liu}\ \emph {et~al.}(2019)\citenamefont {Liu},
  \citenamefont {Li},\ and\ \citenamefont {Li}}]{Liu2019}%
  \BibitemOpen
  \bibfield  {author} {\bibinfo {author} {\bibfnamefont {Q.}~\bibnamefont
  {Liu}}, \bibinfo {author} {\bibfnamefont {H.}~\bibnamefont {Li}}, \ and\
  \bibinfo {author} {\bibfnamefont {M.}~\bibnamefont {Li}},\ }\href {\doibase
  10.1364/OPTICA.6.000778} {\bibfield  {journal} {\bibinfo  {journal} {Optica}\
  }\textbf {\bibinfo {volume} {6}},\ \bibinfo {pages} {778} (\bibinfo {year}
  {2019})}\BibitemShut {NoStop}%
\bibitem [{\citenamefont {Scullion}\ \emph {et~al.}(2015)\citenamefont
  {Scullion}, \citenamefont {Arita}, \citenamefont {Krauss},\ and\
  \citenamefont {Dholakia}}]{scullion2015enhancement}%
  \BibitemOpen
  \bibfield  {author} {\bibinfo {author} {\bibfnamefont {M.~G.}\ \bibnamefont
  {Scullion}}, \bibinfo {author} {\bibfnamefont {Y.}~\bibnamefont {Arita}},
  \bibinfo {author} {\bibfnamefont {T.~F.}\ \bibnamefont {Krauss}}, \ and\
  \bibinfo {author} {\bibfnamefont {K.}~\bibnamefont {Dholakia}},\ }\href@noop
  {} {\bibfield  {journal} {\bibinfo  {journal} {Optica}\ }\textbf {\bibinfo
  {volume} {2}},\ \bibinfo {pages} {816} (\bibinfo {year} {2015})}\BibitemShut
  {NoStop}%
\bibitem [{\citenamefont {Aspelmeyer}\ \emph {et~al.}(2014)\citenamefont
  {Aspelmeyer}, \citenamefont {Kippenberg},\ and\ \citenamefont
  {Marquardt}}]{Aspelmeyer2014}%
  \BibitemOpen
  \bibfield  {author} {\bibinfo {author} {\bibfnamefont {M.}~\bibnamefont
  {Aspelmeyer}}, \bibinfo {author} {\bibfnamefont {T.~J.}\ \bibnamefont
  {Kippenberg}}, \ and\ \bibinfo {author} {\bibfnamefont {F.}~\bibnamefont
  {Marquardt}},\ }\href {\doibase 10.1103/RevModPhys.86.1391} {\bibfield
  {journal} {\bibinfo  {journal} {Reviews of Modern Physics}\ } (\bibinfo
  {year} {2014}),\ 10.1103/RevModPhys.86.1391},\ \Eprint
  {http://arxiv.org/abs/0712.1618} {arXiv:0712.1618} \BibitemShut {NoStop}%
\bibitem [{\citenamefont {Zhang}\ \emph {et~al.}(2012)\citenamefont {Zhang},
  \citenamefont {Wiederhecker}, \citenamefont {Manipatruni}, \citenamefont
  {Barnard}, \citenamefont {McEuen},\ and\ \citenamefont {Lipson}}]{Zhang2012}%
  \BibitemOpen
  \bibfield  {author} {\bibinfo {author} {\bibfnamefont {M.}~\bibnamefont
  {Zhang}}, \bibinfo {author} {\bibfnamefont {G.~S.}\ \bibnamefont
  {Wiederhecker}}, \bibinfo {author} {\bibfnamefont {S.}~\bibnamefont
  {Manipatruni}}, \bibinfo {author} {\bibfnamefont {A.}~\bibnamefont
  {Barnard}}, \bibinfo {author} {\bibfnamefont {P.}~\bibnamefont {McEuen}}, \
  and\ \bibinfo {author} {\bibfnamefont {M.}~\bibnamefont {Lipson}},\ }\href
  {\doibase 10.1103/PhysRevLett.109.233906} {\bibfield  {journal} {\bibinfo
  {journal} {Phys. Rev. Lett.}\ }\textbf {\bibinfo {volume} {109}},\ \bibinfo
  {pages} {233906} (\bibinfo {year} {2012})}\BibitemShut {NoStop}%
\bibitem [{\citenamefont {Dong}\ \emph {et~al.}(2014)\citenamefont {Dong},
  \citenamefont {Zhang}, \citenamefont {Fiore},\ and\ \citenamefont
  {Wang}}]{Dong:14}%
  \BibitemOpen
  \bibfield  {author} {\bibinfo {author} {\bibfnamefont {C.}~\bibnamefont
  {Dong}}, \bibinfo {author} {\bibfnamefont {J.}~\bibnamefont {Zhang}},
  \bibinfo {author} {\bibfnamefont {V.}~\bibnamefont {Fiore}}, \ and\ \bibinfo
  {author} {\bibfnamefont {H.}~\bibnamefont {Wang}},\ }\href {\doibase
  10.1364/OPTICA.1.000425} {\bibfield  {journal} {\bibinfo  {journal} {Optica}\
  }\textbf {\bibinfo {volume} {1}},\ \bibinfo {pages} {425} (\bibinfo {year}
  {2014})}\BibitemShut {NoStop}%
\bibitem [{\citenamefont {Shkarin}\ \emph {et~al.}(2014)\citenamefont
  {Shkarin}, \citenamefont {Flowers-Jacobs}, \citenamefont {Hoch},
  \citenamefont {Kashkanova}, \citenamefont {Deutsch}, \citenamefont
  {Reichel},\ and\ \citenamefont {Harris}}]{ShkarinPRL2014}%
  \BibitemOpen
  \bibfield  {author} {\bibinfo {author} {\bibfnamefont {A.~B.}\ \bibnamefont
  {Shkarin}}, \bibinfo {author} {\bibfnamefont {N.~E.}\ \bibnamefont
  {Flowers-Jacobs}}, \bibinfo {author} {\bibfnamefont {S.~W.}\ \bibnamefont
  {Hoch}}, \bibinfo {author} {\bibfnamefont {A.~D.}\ \bibnamefont
  {Kashkanova}}, \bibinfo {author} {\bibfnamefont {C.}~\bibnamefont {Deutsch}},
  \bibinfo {author} {\bibfnamefont {J.}~\bibnamefont {Reichel}}, \ and\
  \bibinfo {author} {\bibfnamefont {J.~G.~E.}\ \bibnamefont {Harris}},\ }\href
  {\doibase 10.1103/PhysRevLett.112.013602} {\bibfield  {journal} {\bibinfo
  {journal} {Phys. Rev. Lett.}\ }\textbf {\bibinfo {volume} {112}},\ \bibinfo
  {pages} {013602} (\bibinfo {year} {2014})}\BibitemShut {NoStop}%
\bibitem [{\citenamefont {Grutter}\ \emph {et~al.}(2015)\citenamefont
  {Grutter}, \citenamefont {Davan\c{c}o},\ and\ \citenamefont
  {Srinivasan}}]{Grutter:15}%
  \BibitemOpen
  \bibfield  {author} {\bibinfo {author} {\bibfnamefont {K.~E.}\ \bibnamefont
  {Grutter}}, \bibinfo {author} {\bibfnamefont {M.~I.}\ \bibnamefont
  {Davan\c{c}o}}, \ and\ \bibinfo {author} {\bibfnamefont {K.}~\bibnamefont
  {Srinivasan}},\ }\href {\doibase 10.1364/OPTICA.2.000994} {\bibfield
  {journal} {\bibinfo  {journal} {Optica}\ }\textbf {\bibinfo {volume} {2}},\
  \bibinfo {pages} {994} (\bibinfo {year} {2015})}\BibitemShut {NoStop}%
\bibitem [{\citenamefont {Colombano}\ \emph {et~al.}(2019)\citenamefont
  {Colombano}, \citenamefont {Arregui}, \citenamefont {Capuj}, \citenamefont
  {Pitanti}, \citenamefont {Maire}, \citenamefont {Griol}, \citenamefont
  {Garrido}, \citenamefont {Martinez}, \citenamefont {Sotomayor-Torres},\ and\
  \citenamefont {Navarro-Urrios}}]{Colombano2019}%
  \BibitemOpen
  \bibfield  {author} {\bibinfo {author} {\bibfnamefont {M.~F.}\ \bibnamefont
  {Colombano}}, \bibinfo {author} {\bibfnamefont {G.}~\bibnamefont {Arregui}},
  \bibinfo {author} {\bibfnamefont {N.~E.}\ \bibnamefont {Capuj}}, \bibinfo
  {author} {\bibfnamefont {A.}~\bibnamefont {Pitanti}}, \bibinfo {author}
  {\bibfnamefont {J.}~\bibnamefont {Maire}}, \bibinfo {author} {\bibfnamefont
  {A.}~\bibnamefont {Griol}}, \bibinfo {author} {\bibfnamefont
  {B.}~\bibnamefont {Garrido}}, \bibinfo {author} {\bibfnamefont
  {A.}~\bibnamefont {Martinez}}, \bibinfo {author} {\bibfnamefont {C.~M.}\
  \bibnamefont {Sotomayor-Torres}}, \ and\ \bibinfo {author} {\bibfnamefont
  {D.}~\bibnamefont {Navarro-Urrios}},\ }\href {\doibase
  10.1103/PhysRevLett.123.017402} {\bibfield  {journal} {\bibinfo  {journal}
  {Phys. Rev. Lett.}\ }\textbf {\bibinfo {volume} {123}},\ \bibinfo {pages}
  {017402} (\bibinfo {year} {2019})}\BibitemShut {NoStop}%
\bibitem [{\citenamefont {Djorwe}\ \emph {et~al.}(2018)\citenamefont {Djorwe},
  \citenamefont {Pennec},\ and\ \citenamefont {Djafari-Rouhani}}]{PREDjorwe}%
  \BibitemOpen
  \bibfield  {author} {\bibinfo {author} {\bibfnamefont {P.}~\bibnamefont
  {Djorwe}}, \bibinfo {author} {\bibfnamefont {Y.}~\bibnamefont {Pennec}}, \
  and\ \bibinfo {author} {\bibfnamefont {B.}~\bibnamefont {Djafari-Rouhani}},\
  }\href {\doibase 10.1103/PhysRevE.98.032201} {\bibfield  {journal} {\bibinfo
  {journal} {Phys. Rev. E}\ }\textbf {\bibinfo {volume} {98}},\ \bibinfo
  {pages} {032201} (\bibinfo {year} {2018})}\BibitemShut {NoStop}%
\bibitem [{\citenamefont {del Pino}\ \emph {et~al.}(2022)\citenamefont {del
  Pino}, \citenamefont {Slim},\ and\ \citenamefont {Verhagen}}]{delPino2022}%
  \BibitemOpen
  \bibfield  {author} {\bibinfo {author} {\bibfnamefont {J.}~\bibnamefont {del
  Pino}}, \bibinfo {author} {\bibfnamefont {J.~J.}\ \bibnamefont {Slim}}, \
  and\ \bibinfo {author} {\bibfnamefont {E.}~\bibnamefont {Verhagen}},\ }\href
  {\doibase 10.1038/s41586-022-04609-0} {\bibfield  {journal} {\bibinfo
  {journal} {Nature}\ }\textbf {\bibinfo {volume} {606}},\ \bibinfo {pages}
  {82} (\bibinfo {year} {2022})}\BibitemShut {NoStop}%
\bibitem [{\citenamefont {Ruesink}\ \emph {et~al.}(2018)\citenamefont
  {Ruesink}, \citenamefont {Mathew}, \citenamefont {Miri}, \citenamefont
  {Al{\`u}},\ and\ \citenamefont {Verhagen}}]{ruesink2018optical}%
  \BibitemOpen
  \bibfield  {author} {\bibinfo {author} {\bibfnamefont {F.}~\bibnamefont
  {Ruesink}}, \bibinfo {author} {\bibfnamefont {J.~P.}\ \bibnamefont {Mathew}},
  \bibinfo {author} {\bibfnamefont {M.-A.}\ \bibnamefont {Miri}}, \bibinfo
  {author} {\bibfnamefont {A.}~\bibnamefont {Al{\`u}}}, \ and\ \bibinfo
  {author} {\bibfnamefont {E.}~\bibnamefont {Verhagen}},\ }\href@noop {}
  {\bibfield  {journal} {\bibinfo  {journal} {Nature communications}\ }\textbf
  {\bibinfo {volume} {9}},\ \bibinfo {pages} {1} (\bibinfo {year}
  {2018})}\BibitemShut {NoStop}%
\bibitem [{\citenamefont {De~L{\'e}pinay}\ \emph {et~al.}(2020)\citenamefont
  {De~L{\'e}pinay}, \citenamefont {Ockeloen-Korppi}, \citenamefont {Malz},\
  and\ \citenamefont {Sillanp{\"a}{\"a}}}]{de2020nonreciprocal}%
  \BibitemOpen
  \bibfield  {author} {\bibinfo {author} {\bibfnamefont {L.~M.}\ \bibnamefont
  {De~L{\'e}pinay}}, \bibinfo {author} {\bibfnamefont {C.~F.}\ \bibnamefont
  {Ockeloen-Korppi}}, \bibinfo {author} {\bibfnamefont {D.}~\bibnamefont
  {Malz}}, \ and\ \bibinfo {author} {\bibfnamefont {M.~A.}\ \bibnamefont
  {Sillanp{\"a}{\"a}}},\ }\href@noop {} {\bibfield  {journal} {\bibinfo
  {journal} {Physical Review Letters}\ }\textbf {\bibinfo {volume} {125}},\
  \bibinfo {pages} {023603} (\bibinfo {year} {2020})}\BibitemShut {NoStop}%
\bibitem [{\citenamefont {Albrechtsen}\ \emph {et~al.}(2021)\citenamefont
  {Albrechtsen}, \citenamefont {Lahijani}, \citenamefont {Christiansen},
  \citenamefont {Nguyen}, \citenamefont {Casses}, \citenamefont {Hansen},
  \citenamefont {Stenger}, \citenamefont {Sigmund}, \citenamefont {Jansen},
  \citenamefont {Mørk},\ and\ \citenamefont {Stobbe}}]{Albrechtsen2021arXiv}%
  \BibitemOpen
  \bibfield  {author} {\bibinfo {author} {\bibfnamefont {M.}~\bibnamefont
  {Albrechtsen}}, \bibinfo {author} {\bibfnamefont {B.~V.}\ \bibnamefont
  {Lahijani}}, \bibinfo {author} {\bibfnamefont {R.~E.}\ \bibnamefont
  {Christiansen}}, \bibinfo {author} {\bibfnamefont {V.~T.~H.}\ \bibnamefont
  {Nguyen}}, \bibinfo {author} {\bibfnamefont {L.~N.}\ \bibnamefont {Casses}},
  \bibinfo {author} {\bibfnamefont {S.~E.}\ \bibnamefont {Hansen}}, \bibinfo
  {author} {\bibfnamefont {N.}~\bibnamefont {Stenger}}, \bibinfo {author}
  {\bibfnamefont {O.}~\bibnamefont {Sigmund}}, \bibinfo {author} {\bibfnamefont
  {H.}~\bibnamefont {Jansen}}, \bibinfo {author} {\bibfnamefont
  {J.}~\bibnamefont {Mørk}}, \ and\ \bibinfo {author} {\bibfnamefont
  {S.}~\bibnamefont {Stobbe}},\ }\href@noop {} {\enquote {\bibinfo {title}
  {Nanometer-scale photon confinement inside dielectrics},}\ } (\bibinfo {year}
  {2021}),\ \Eprint {http://arxiv.org/abs/2108.01681} {arXiv:2108.01681
  [physics.optics]} \BibitemShut {NoStop}%
\bibitem [{\citenamefont {Nguyen}\ \emph {et~al.}(2020)\citenamefont {Nguyen},
  \citenamefont {Silvestre}, \citenamefont {Shi}, \citenamefont {Cork},
  \citenamefont {Jensen}, \citenamefont {Hubner}, \citenamefont {Ma},
  \citenamefont {Leussink}, \citenamefont {de~Boer},\ and\ \citenamefont
  {Jansen}}]{Nguyen2020}%
  \BibitemOpen
  \bibfield  {author} {\bibinfo {author} {\bibfnamefont {V.~T.~H.}\
  \bibnamefont {Nguyen}}, \bibinfo {author} {\bibfnamefont {C.}~\bibnamefont
  {Silvestre}}, \bibinfo {author} {\bibfnamefont {P.}~\bibnamefont {Shi}},
  \bibinfo {author} {\bibfnamefont {R.}~\bibnamefont {Cork}}, \bibinfo {author}
  {\bibfnamefont {F.}~\bibnamefont {Jensen}}, \bibinfo {author} {\bibfnamefont
  {J.}~\bibnamefont {Hubner}}, \bibinfo {author} {\bibfnamefont
  {K.}~\bibnamefont {Ma}}, \bibinfo {author} {\bibfnamefont {P.}~\bibnamefont
  {Leussink}}, \bibinfo {author} {\bibfnamefont {M.}~\bibnamefont {de~Boer}}, \
  and\ \bibinfo {author} {\bibfnamefont {H.}~\bibnamefont {Jansen}},\ }\href
  {\doibase 10.1149/2162-8777/ab61ed} {\bibfield  {journal} {\bibinfo
  {journal} {{ECS} Journal of Solid State Science and Technology}\ }\textbf
  {\bibinfo {volume} {9}},\ \bibinfo {pages} {024002} (\bibinfo {year}
  {2020})}\BibitemShut {NoStop}%
\bibitem [{\citenamefont {Hoang~Nguyen}\ \emph {et~al.}(2021)\citenamefont
  {Hoang~Nguyen}, \citenamefont {Jensen}, \citenamefont {Hübner},
  \citenamefont {Shkondin}, \citenamefont {Cork}, \citenamefont {Ma},
  \citenamefont {Leussink}, \citenamefont {De~Malsche},\ and\ \citenamefont
  {Jansen}}]{Nguyen2021}%
  \BibitemOpen
  \bibfield  {author} {\bibinfo {author} {\bibfnamefont {V.~T.}\ \bibnamefont
  {Hoang~Nguyen}}, \bibinfo {author} {\bibfnamefont {F.}~\bibnamefont
  {Jensen}}, \bibinfo {author} {\bibfnamefont {J.}~\bibnamefont {Hübner}},
  \bibinfo {author} {\bibfnamefont {E.}~\bibnamefont {Shkondin}}, \bibinfo
  {author} {\bibfnamefont {R.}~\bibnamefont {Cork}}, \bibinfo {author}
  {\bibfnamefont {K.}~\bibnamefont {Ma}}, \bibinfo {author} {\bibfnamefont
  {P.}~\bibnamefont {Leussink}}, \bibinfo {author} {\bibfnamefont
  {W.}~\bibnamefont {De~Malsche}}, \ and\ \bibinfo {author} {\bibfnamefont
  {H.}~\bibnamefont {Jansen}},\ }\href {\doibase 10.1116/6.0000922} {\bibfield
  {journal} {\bibinfo  {journal} {Journal of Vacuum Science \& Technology B}\
  }\textbf {\bibinfo {volume} {39}},\ \bibinfo {pages} {032201} (\bibinfo
  {year} {2021})},\ \Eprint
  {http://arxiv.org/abs/https://doi.org/10.1116/60000922}
  {https://doi.org/10.1116/60000922} \BibitemShut {NoStop}%
\bibitem [{\citenamefont {Navarro-Urrios}\ \emph {et~al.}(2014)\citenamefont
  {Navarro-Urrios}, \citenamefont {Gomis-Bresco}, \citenamefont {Capuj},
  \citenamefont {Alzina}, \citenamefont {Griol}, \citenamefont {Puerto},
  \citenamefont {MartÃ­nez},\ and\ \citenamefont
  {Sotomayor-Torres}}]{Navarro-Urrios2014}%
  \BibitemOpen
  \bibfield  {author} {\bibinfo {author} {\bibfnamefont {D.}~\bibnamefont
  {Navarro-Urrios}}, \bibinfo {author} {\bibfnamefont {J.}~\bibnamefont
  {Gomis-Bresco}}, \bibinfo {author} {\bibfnamefont {N.~E.}\ \bibnamefont
  {Capuj}}, \bibinfo {author} {\bibfnamefont {F.}~\bibnamefont {Alzina}},
  \bibinfo {author} {\bibfnamefont {A.}~\bibnamefont {Griol}}, \bibinfo
  {author} {\bibfnamefont {D.}~\bibnamefont {Puerto}}, \bibinfo {author}
  {\bibfnamefont {A.}~\bibnamefont {MartÃ­nez}}, \ and\ \bibinfo {author}
  {\bibfnamefont {C.~M.}\ \bibnamefont {Sotomayor-Torres}},\ }\href {\doibase
  10.1063/1.4894623} {\bibfield  {journal} {\bibinfo  {journal} {Journal of
  Applied Physics}\ }\textbf {\bibinfo {volume} {116}},\ \bibinfo {pages}
  {093506} (\bibinfo {year} {2014})},\ \Eprint
  {http://arxiv.org/abs/https://doi.org/10.1063/1.4894623}
  {https://doi.org/10.1063/1.4894623} \BibitemShut {NoStop}%
\bibitem [{\citenamefont {Maire}\ \emph {et~al.}(2018)\citenamefont {Maire},
  \citenamefont {Arregui}, \citenamefont {Capuj}, \citenamefont {Colombano},
  \citenamefont {Griol}, \citenamefont {Martinez}, \citenamefont
  {Sotomayor-Torres},\ and\ \citenamefont {Navarro-Urrios}}]{Maire2018}%
  \BibitemOpen
  \bibfield  {author} {\bibinfo {author} {\bibfnamefont {J.}~\bibnamefont
  {Maire}}, \bibinfo {author} {\bibfnamefont {G.}~\bibnamefont {Arregui}},
  \bibinfo {author} {\bibfnamefont {N.~E.}\ \bibnamefont {Capuj}}, \bibinfo
  {author} {\bibfnamefont {M.~F.}\ \bibnamefont {Colombano}}, \bibinfo {author}
  {\bibfnamefont {A.}~\bibnamefont {Griol}}, \bibinfo {author} {\bibfnamefont
  {A.}~\bibnamefont {Martinez}}, \bibinfo {author} {\bibfnamefont {C.~M.}\
  \bibnamefont {Sotomayor-Torres}}, \ and\ \bibinfo {author} {\bibfnamefont
  {D.}~\bibnamefont {Navarro-Urrios}},\ }\href {\doibase 10.1063/1.5040061}
  {\bibfield  {journal} {\bibinfo  {journal} {APL Photonics}\ }\textbf
  {\bibinfo {volume} {3}},\ \bibinfo {pages} {126102} (\bibinfo {year}
  {2018})},\ \Eprint {http://arxiv.org/abs/https://doi.org/10.1063/1.5040061}
  {https://doi.org/10.1063/1.5040061} \BibitemShut {NoStop}%
\bibitem [{\citenamefont {Gorodetsky}\ \emph {et~al.}(2010)\citenamefont
  {Gorodetsky}, \citenamefont {Schliesser}, \citenamefont {Anetsberger},
  \citenamefont {Deleglise},\ and\ \citenamefont
  {Kippenberg}}]{Gorodetsky2010}%
  \BibitemOpen
  \bibfield  {author} {\bibinfo {author} {\bibfnamefont {M.}~\bibnamefont
  {Gorodetsky}}, \bibinfo {author} {\bibfnamefont {A.}~\bibnamefont
  {Schliesser}}, \bibinfo {author} {\bibfnamefont {G.}~\bibnamefont
  {Anetsberger}}, \bibinfo {author} {\bibfnamefont {S.}~\bibnamefont
  {Deleglise}}, \ and\ \bibinfo {author} {\bibfnamefont {T.}~\bibnamefont
  {Kippenberg}},\ }\href {\doibase 10.1364/OE.18.023236} {\bibfield  {journal}
  {\bibinfo  {journal} {Optics express}\ }\textbf {\bibinfo {volume} {18}},\
  \bibinfo {pages} {23236} (\bibinfo {year} {2010})}\BibitemShut {NoStop}%
\end{thebibliography}%

\end{document}